\documentclass[lettersize,journal]{IEEEtran}
\usepackage{amsmath,amsfonts}
\usepackage{algorithmic}
\usepackage{algorithm}
\usepackage{array}
\usepackage[caption=false,font=normalsize,labelfont=sf,textfont=sf]{subfig}
\usepackage{textcomp}
\usepackage{stfloats}
\usepackage{url}
\usepackage{verbatim}
\usepackage{graphicx}
\usepackage{cite}
\usepackage{multirow}
\usepackage{makecell}
\begin{document}

\title{A Methodology for Power Dispatch Based on Traction Station Clusters in the Flexible Traction Power Supply System}

\author{Ruofan~Li,
	Qianhao~Sun,
	Qifang~Chen,
	and~Mingchao~Xia,
	\thanks{This work was supported by the Key Program of the National Natural Science Foundation of China (Grant No. 52337003) and the Young Scientists Fund of the National Natural Science Foundation of China (Grant No. 52307210). \textit{(Corresponding author: Mingchao Xia.)}}
	\thanks{Ruofan Li, Qianhao Sun and Qifang Chen are with the School of Electrical engineering, Beijing Jiaotong University, Beijing 100044, China (e-mail: 23111461@bjtu.edu.cn; sunqianhao@bjtu.edu.cn; chenqf@bjtu.edu.cn).}
	\thanks{Mingchao Xia is with the School of Electrical engineering, Beijing Jiaotong University, Beijing 100044, China and also with the School of Electrical Engineering, Shenyang University of Technology, Shenyang 110870, China (e-mail: mchxia@bjtu.edu.cn).}
}

\markboth{IEEE TRANSACTIONS ON TRANSPORTATION ELECTRIFICATION}%
{Shell \MakeLowercase{\textit{et al.}}: A Sample Article Using IEEEtran.cls for IEEE Journals}


\maketitle
\pagestyle{empty}
\thispagestyle{empty}

\begin{abstract}
The flexible traction power supply system (FTPSS) eliminates the neutral zone but leads to increased complexity in power flow coordinated control and power mismatch. To address these challenges, the methodology for power dispatch (PD) based on traction station clusters (TSCs) in FTPSS is proposed, in which each TSC with a consistent structure performs independent local phase angle control. First, to simplify the PD problem of TSCs, the system is transformed into an equivalent model with constant topology, resulting in it can be solved by univariate numerical optimization with higher computational performance. Next, the calculation method of the feasible phase angle domain under strict and relaxed power circulation constraints are described, respectively, which ensures that power circulation can be either eliminated or precisely controlled. Finally, the PD method with three unique modes for uncertain train loads is introduced to enhance power flow flexibility: specified power distribution coefficients between traction substations (TSs), constant output power of TSs, and maximum consumption of renewable resources within TSs. In the experimental section, the performance of the TSC methodology for PD is verified through detailed train operation scenarios.
\end{abstract}

\begin{IEEEkeywords}
Flexible traction power supply system (FTPSS), power flow control, power circulation, energy management.
\end{IEEEkeywords}

\section{Introduction}
\IEEEPARstart{I}{n} recent years, with the continuous expansion of electrified railway operational mileage, the supply-demand imbalance between railway systems and power systems has become increasingly acute. Renewable energy sources (RES) have broad application prospects in the railway sector, providing a solution to this issue \cite{chenVoltageUnbalanceProbability2023, cheng_solar-powered_2022}. In the context of high RES penetration  and the secondary utilization of train (TRs) regenerative braking energy (BRE), enhancing the performance of power flow control for the traction power supply system (TPSS) is of great significance for promoting low-carbon operation and eliminating the neutral zone (NZ).

Extensive literature has been reported on traditional TPSS regarding the integrated energy management needs for BRE and RES. In traditional TPSS, power flow controllers (PFCs) are installed at the exits of traction substations (TSs) and the ends of power supply sections to actively control power flows caused by TR loads and BRE \cite{yuan_optimal_2022}. To enhance the utilization of BRE and RES energy, various strategies have been proposed, including rolling adaptive peak clipping control \cite{ma_rolling-adaptive_2023}, model predictive control \cite{ge_combined_2022,novak_hierarchical_2019}, hierarchical coordinated control \cite{chen_integrated_2023,ge_hierarchical_2023} and multiple optimization strategies \cite{huang_joint_2024, wangPowerAllocationStrategy2023}. Additionally, various compensation methods have been proposed to improve power quality \cite{zhang_analysis_2023, feng_evaluation_2020, zahedmanesh_sequential_2021, lin_energy_2023}. However, traditional TPSS struggle to eliminate NZs that pose operational risks to TR. Consequently, the co-phase traction power supply system (CTPSS) and flexible traction power supply system (FTPSS) have gradually become a research hotspot.

In CTPSS, the TS consists of a single-phase transformer and a parallel PFC operating in PQ control mode \cite{chen_bi-hierarchy_2023}. Various operating modes of CTPSS have been explored \cite{chen_configuration_2023}, and methods such as chance-constrained optimization \cite{chen_chance-constrained_2024} and two-stage robust optimization \cite{liu_robust_2022} have been proposed to enhance the system's ability to cope with uncertain PV and traction loads.
However, CTPSS also exposes some issues. Literature \cite{zhang_research_2022} indicates that differences in output voltage between adjacent TS can lead to power circulation and high electricity costs, which require specialized technical solutions. Additionally, the negative sequence current on the grid side is caused by both the transformer and the PFC's active power \cite{liu_robust_2022}. The additional power compensation results in increased constraints and more complex control strategies \cite{chen_dynamic_2022, chen_unified_2022, chen_modelling_2021, liu_co-phase_2020}. 

With the development of high-power power electronics technology, FTPSS schemes based on AC-DC-AC converters have been progressively explored. Several studies have been conducted on energy management for FTPSS, including reinforcement learning methods \cite{ying_online_2023}, mixed-integer programming methods \cite{chen_optimal_2022}, and day-ahead and intra-day scheduling methods combined with model predictive control \cite{chen_multitime-scale_2020}. Additionally, to achieve controllable distribution of traction loads to multiple TS, a few studies have explored droop control. For instance, literature \cite{li_modeling_2023} proposed a power sharing control method based on virtual power decoupling. However, although FTPSS eliminates NZs and facilitates the integrated management of power flows and BRE through flexible power channels between TSs, the direct electrical coupling of TSs introduces several issues that have not yet been fully addressed: 1) Power flow scheduling requires global coordinated control of multiple TSs, necessitating long-distance transmission of TR operational information and TS control instructions. The inevitable signal transmission delay poses challenges to the real-time coordinated control. Although a few studies have reported droop control schemes aimed at achieving distributed local control and controlled power allocation of TS \cite{li_modeling_2023}. However, it leads to unsafe long-distance power transmission across power supply zones and higher energy losses \cite{vazquezFullyDecentralizedAdaptive2019}. 2) In FTPSS, the cooperative controller calculates the operation instructions of each TS in real time according to the status of TR and RES \cite{ying_online_2023, chen_multitime-scale_2020}. However, the uncertain TR loads and online calculation delay cause instantaneous mismatch between TS output power and traction demand, which leads to potential frequency and voltage fluctuations in the traction network. 3) Voltage and phase angle mismatches between adjacent TSs  lead to significant power circulation \cite{zhang_research_2022}. For power electronic converters with limited capacity, accurate management and suppression of power circulating currents is crucial. To this end, a methodology for power dispatch (PD) based on traction substation clusters (TSCs) is proposed. The primary contributions are as follows:

\begin{enumerate}
	\item{To address the challenges of real-time coordinated control and instantaneous power mismatches, a TSC scheme in FTPSS has been proposed. TSs are divided into multiple independent TSCs that perform local phase angle control internally. This scheme eliminates the need for global power flow scheduling and provides the stable power channel for uncertain TR loads. Compared to droop control, the power isolation characteristics of TSCs avoid long-distance power transmission.}
	\item{To address the challenges of power circulation between TSs and online computation delays, the feasible phase angle domain (FPAD) under strict active and reactive power circulation constraints, as well as relaxed constraints, is thoroughly discussed along with its calculation method. Based on an equivalent modeling method for FTPSS, the original system is transformed into a standard topology, allowing the algorithm to perform single-variable numerical optimization with higher computational performance.}
	\item{Based on the equivalent model, the flexible PD method for FTPSS is proposed, which enables the system to support three switchable PD modes: the inter-station power distribution mode (PDM), the constant power output mode (CPM) and the RES maximum consumption mode (MCM). These modes significantly enhance the flexibility of FTPSS in managing RES and BRE.}
\end{enumerate}

\section{Proposed FTPSS with TCSs}
\subsection{System Description}
Fig. 1 (a) illustrates the proposed FTPSS with TSCs based on phase angle control.
In the system, each TS is composed of an AC-DC-AC power electronic converter and maintains the same output voltage magnitude, ${{U}_{\text{N}}}$. RES and ESS are uniformly connected to the DC bus within the converter. Based on the global positioning system (GPS) unified timing, phase measurement units (PMUs) provide each TS with a synchronized phase, $\theta_{\text{sync}}$ \cite{qiuPulsarCalibratedTimingSource2022, phadke_phasor_2018}. The system alternates between two types of TSs along the traction network. One type is the N-TS, which maintains synchronized output voltage phases based on $\theta_{\text{sync}}$ and a constant frequency of 50 Hz. All N-TSs output voltage is ${{U}_{\text{N}}}\angle{{0}^{\circ}}$. The other type is the A-TS, with a flexibly adjustable phase. The real-time phase angle controller (RPAC) calculates the reference phase angle (RPA) ${{\delta}_{i}}$ for each A-TS based on the system's status, where $i = a, b, c, \ldots, {{N}_{\text{A-TS}}}$, ${{N}_{\text{A-TS}}}$ is the number of A-TSs in the system. Thus, the voltage of the A-TSs is ${{U}_{\text{N}}}\angle {{\delta }_{i}}$.

Two adjacent N-TSs and one intermediate A-TS form a TSC. Each TSC contains a double-track traction network connected in parallel, with an uncertain number of TRs. As indicated by the red dashed lines in the figure, the structure composed of the TSs on both sides and a single traction network is called the minimum supply organization (MSO). Two MSOs connected in parallel form a zone supply organization (ZSO). The complete FTPSS can be viewed as a continuous extension of several identical TSCs. Since all TS's converters operate in the $\text{V}/\theta$ control mode, this ensures that even with TR load fluctuations, the TS output power can automatically balance with the traction load. Additionally, the presence of N-TSs ensures the power decoupling between two adjacent TSCs. Adjusting the phase angle of the A-TS within one TSC does not affect the power flow in the neighboring TSC. Therefore, compared to the global power flow scheduling in traditional FTPSS, each TSC performs independent local control. This approach avoids additional interactions between TSCs and allows each RPAC to only acquire the TR operational information within its respective TSC, reducing communication pressure.
\begin{figure}[!t]
	\centering
	\subfloat[]{\includegraphics[width=3.4in]{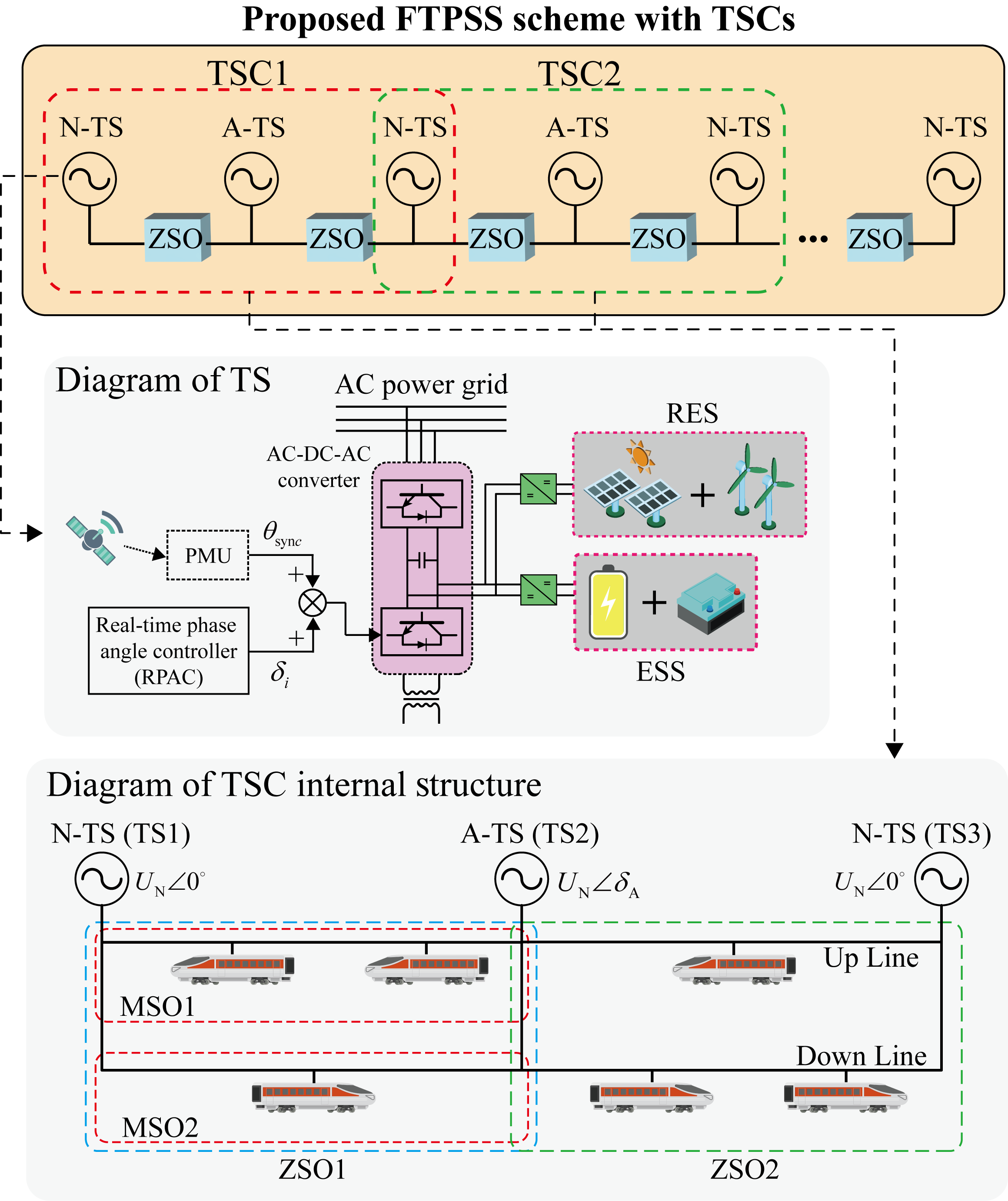}}
	\hfil
	\subfloat[]{\includegraphics[width=3.4in]{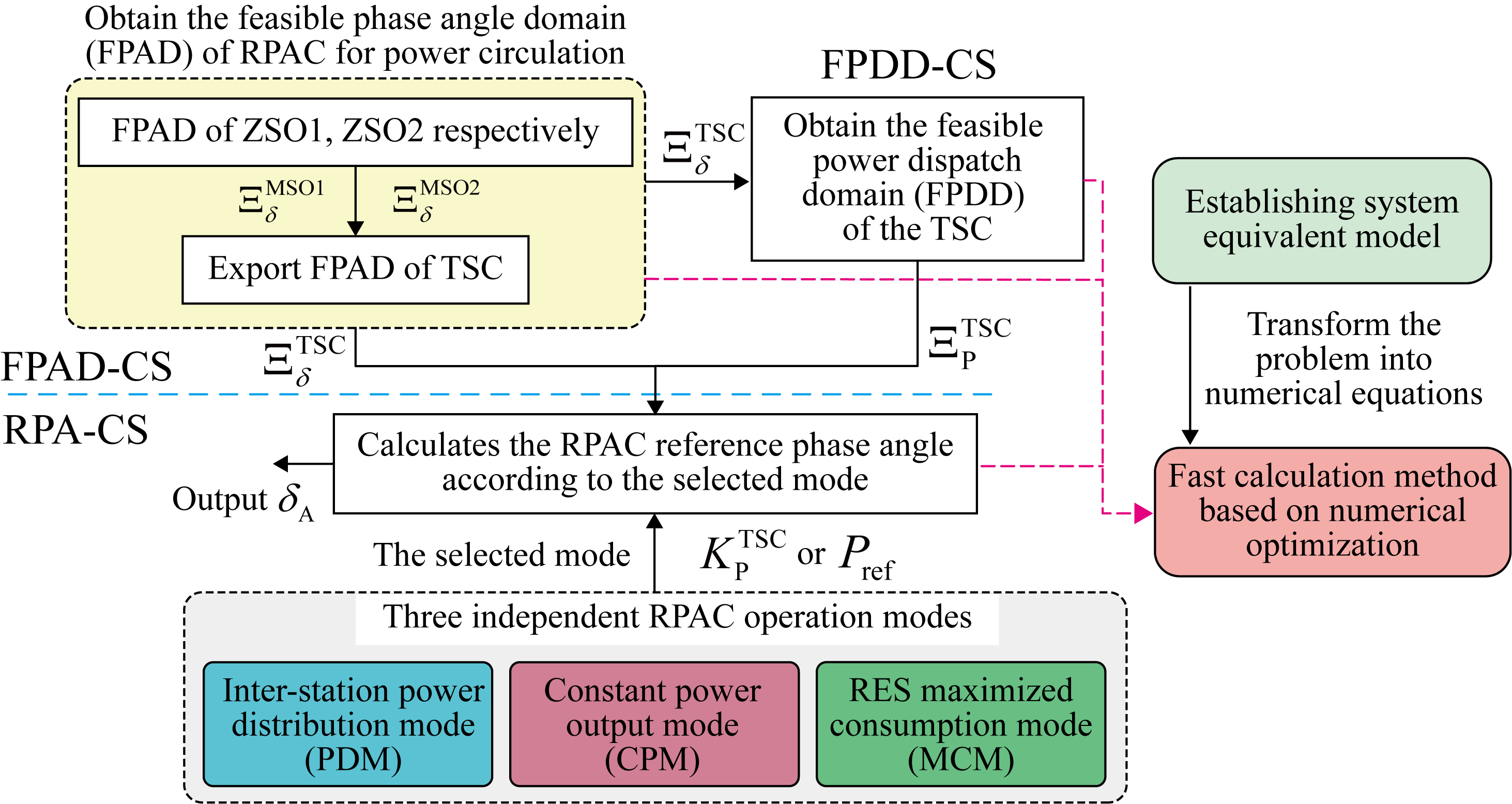}}
	\caption{Proposed FTPSS. (a) the FTPSS scheme with TSCs based on phase angle control. (b) Flowchart of system operation.}
\end{figure}

\subsection{Power Flow Management Solutions}
Fig. 1 (b) shows the flowchart of RPAC performing phase angle control. Since phase angle adjustment affects the power flow within the TSC, the RPAC first obtains FPAD for the A-TS based on the system operating status. During the FPAD calculation stage (FPAD-CS), the active and reactive power circulation within the TSC can be completely suppressed or precisely controlled. In the next RPA calculation stage (RPA-CS), the RPAC calculates the desired operating phase angle for the A-TS in real time to manage the power flow within the TSC. Each TSC in the proposed FTPSS independently supports three flexible PD modes: PDM, CPM, and MCM. In PDM, the output power of the A-TS and N-TS within the TSC is precisely distributed, maintaining balance despite varying TR numbers and loads, similar to droop control characteristics. However, the power decoupling between TSCs prevents the long-distance power transmission issues that may occur with droop schemes. In CPM, the RPAC simulates the traditional FTPSS converter's PQ control strategy by adjusting the phase angle, allowing the A-TS to output a given reference power according to dispatch instructions. In traditional FTPSS, the BRE of trains returns to the TS and is stored by the ESS, which is not beneficial for TSs with abundant RES. To address this, in MCM, the RPAC maximizes A-TS power output and minimizes BRE recovery, with the strategy strictly confined within the PD limits to avoid unexpected power circulation. Moreover, the structural similarity of TSCs allows the established system equivalence model to provide a fast problem-solving method for FTPSS. This ensures that the RPAC does not need to perform any online power flow calculations during the FPAD-CS, RPA-CS and feasible power dispatch domain (FPDD) calculation stage (FPDD-CS).

\section{Method for Obtaining the FPAD}
\subsection{System Equivalent Model}
\begin{figure}[!t]
	\centering
	\subfloat[]{\includegraphics[width=3.4in]{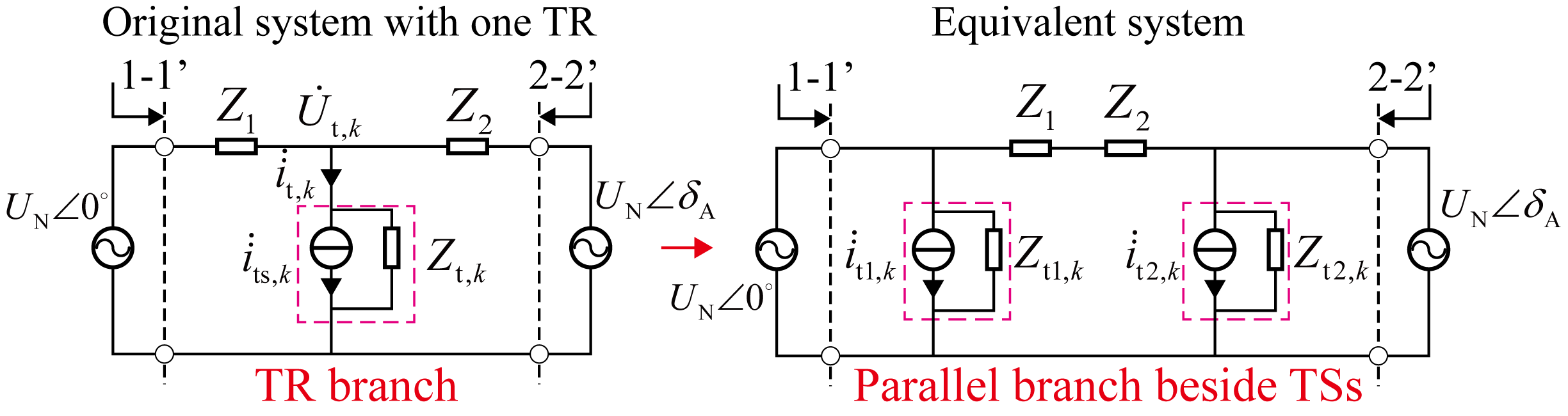}}
	\hfil
	\subfloat[]{\includegraphics[width=2.3in]{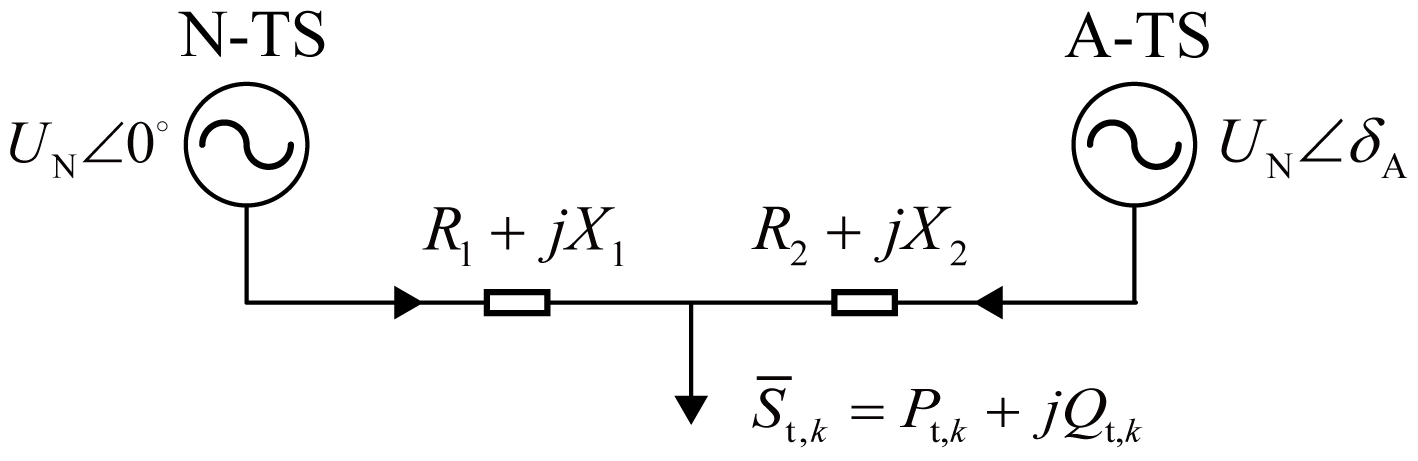}}
	\hfil
	\subfloat[]{\includegraphics[width=2.8in]{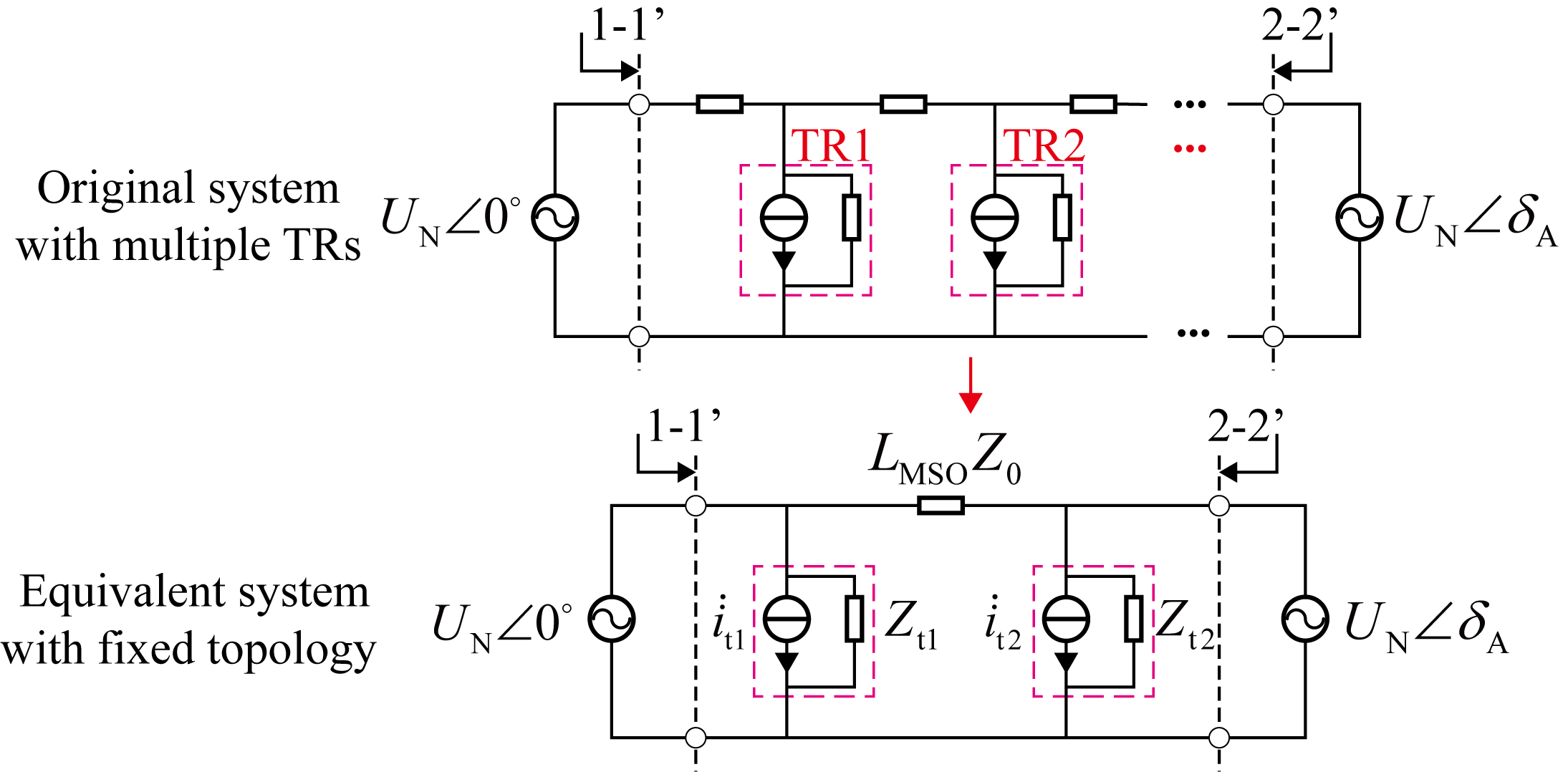}}
	\caption{System equivalent mothod. (a) Equivalent model for a TR branch. (b) Resistance-inductance model of bilateral power supply by N-TS and A-TS. (c) Equivalent method with $N_{\text{TR}}$ TR branches.}
\end{figure}

Fig. 2(a) shows the equivalent FTPSS system containing a TR branch $k$. The TR branch is transformed into a parallel branch with a current source and impedance placed next to TS1 and TS2 on either side. The original system and the equivalent system should ensure the same electrical characteristics at ports 1-1’ and 2-2’. In the parallel branch of the TS, the analytical expressions for the current sources ${\dot{i}}_{\text{t}1,k}$, ${\dot{i}}_{\text{t2},k}$ and the impedances ${Z}_{\text{t1},k}$, ${Z}_{\text{t2},k}$ are:
\begin{equation}
	\begin{cases}
		{{{\dot{i}}}_{\text{t}1,k}}=\frac{1}{{{Z}_{1}}+{{Z}_{2}}}{{U}_{\text{N}}}\angle {{\delta }_{\text{A}}}+\frac{{{Z}_{2}}{{Z}_{\text{t},k}}}{{{Z}_{1}}{{Z}_{t}}+{{Z}_{2}}{{Z}_{\text{t},k}}+{{Z}_{1}}{{Z}_{2}}}\left( {{{\dot{i}}}_{\text{ts},k}}-\frac{{{U}_{\text{N}}}\angle {{\delta }_{\text{A}}}}{{{Z}_{2}}} \right)  \\
		{{Z}_{\text{t}1,k}}=\frac{\left( {{Z}_{1}}+{{\text{Z}}_{2}} \right)\left( {{Z}_{1}}{{Z}_{\text{t},k}}+{{Z}_{2}}{{Z}_{\text{t},k}}+{{Z}_{1}}{{Z}_{2}} \right)}{Z_{2}^{2}}  \\
	\end{cases}
\end{equation}
\begin{equation}
	\begin{cases}
		{{{\dot{i}}}_{\text{t2},k}}=\frac{1}{{{Z}_{1}}+{{Z}_{2}}}{{U}_{\text{N}}}\angle {{0}^{{}^\circ }}+\frac{{{Z}_{1}}{{Z}_{\text{t},k}}}{{{Z}_{1}}{{Z}_{t}}+{{Z}_{2}}{{Z}_{\text{t},k}}+{{Z}_{1}}{{Z}_{2}}}\left( {{{\dot{i}}}_{\text{ts},k}}-\frac{{{U}_{\text{N}}}\angle {{0}^{{}^\circ }}}{{{Z}_{1}}} \right)  \\
		{{Z}_{\text{t2},k}}=\frac{\left( {{Z}_{1}}+{{Z}_{2}} \right)\left( {{Z}_{1}}{{Z}_{\text{t},k}}+{{Z}_{2}}{{Z}_{\text{t},k}}+{{Z}_{1}}{{Z}_{2}} \right)}{Z_{1}^{2}}  \\
	\end{cases}
\end{equation}
where ${{Z}_{\text{t},k}}$ represents the TR equivalent impedance; ${{U}_{\text{N}}}\angle {{0}^\circ}$ and ${{U}_{\text{N}}}\angle {{\delta}_{\text{A}}}$ are the output voltage vectors of TS1 and TS2, respectively; ${{Z}_{1}}$ and ${{Z}_{2}}$ are the traction network impedances on either side of the TR branch; ${{\dot{i}}_{\text{ts},k}}$ represents the TR equivalent current source, which can be further expressed as:
\begin{equation}
	{{\dot{i}}_{\text{ts},k}}={{\left( \frac{{{{\bar{S}}}_{\text{t},k}}}{{{{\dot{U}}}_{\text{t},k}}} \right)}^{*}}-\frac{{{{\dot{U}}}_{\text{t},k}}}{{{Z}_{\text{t},k}}}
\end{equation}
where ${{\dot{U}}_{\text{t},k}}$ represents the voltage of the traction network at the TR's pantograph; ${{\bar{S}}_{\text{t},k}}={{P}_{\text{t},k}}+j{{Q}_{\text{t},k}}$ represents the complex power of the TR. During the system equivalence process, the constancy of ${{\bar{S}}_{\text{t},k}}$ must be ensured. Therefore, ${{\dot{i}}_{\text{ts},k}}$ can be considered as a current source controlled by ${{\dot{U}}_{\text{t},k}}$. However, as ${{\dot{U}}_{\text{t},k}}$ is a free variable, an additional equation is needed to make the system of equations solvable. Fig. 2(b) shows a bilateral power supply system containing N-TS and A-TS. Considering that the impedance per unit length of the traction network can be regarded as a constant value (for this paper, the typical value is set as ${{Z}_{0}}=\left(0.15+j 0.55\right)\Omega$), the resistances ${{R}_{1}}$, ${{R}_{2}}$ and the reactances ${{X}_{1}}$, ${{X}_{2}}$ can be converted into constants relative to the TR's position. The analytical expressions for the voltage amplitude ${{U}_{\text{t},k}}$ and phase angle ${{\delta }_{\text{t},k}}$ of ${{\dot{U}}_{\text{t},k}}$ are as follows:
\begin{equation}
	\begin{cases}
		{{\delta }_{\text{t,}k}}=\frac{{{L}_{1}}}{L_{\text{MSO}}}{{\delta }_{\text{A}}}-\frac{a}{b}\left( 	\frac{1}{2}-\frac{1}{2\sqrt{5}}c \right)  \\
		{{U}_{\text{t,}k}}=\left( \frac{1}{2}+\frac{1}{2\sqrt{5}}c \right){{U}_{\text{N}}}  \\
		a=11{{P}_{\text{t,}k}}-3{{Q}_{\text{t,}k}}  \\
		b=3{{P}_{\text{t,}k}}+11{{Q}_{\text{t,}k}}  \\
		c=\sqrt{5-\frac{{{L}_{1}}{{L}_{2}}}{L_{\text{MSO}}U_{\text{N}}^{2}}b}  \\
	\end{cases}
\end{equation}
where ${{L}_{1}}$ and ${{L}_{2}}$ are distances from the TR to TS1 and TS2, respectively; ${{L}_{\text{MSO}}}={{L}_{1}}+{{L}_{2}}$ is the total length of the MSO. Based on Eq. (4), the parallel branch parameters of the TS for TR branch $k$ can be expressed as a numerical equation with respect to ${{\delta}_{\text{A}}}$.

Fig. 2(c) illustrates the system equivalence method for a system containing $N_{\text{TR}}$ TR branches. Since ${{Z}_{\text{t},k}}$ is much larger than the traction network impedance, the original system can be considered as the superposition of multiple subsystems, each containing a single TR branch. The parallel impedances at the TS, ${{Z}_{\text{t1}}}$ and ${{Z}_{\text{t2}}}$, and the current sources ${{\dot{i}}_{\text{t1}}}$ and ${{\dot{i}}_{\text{t2}}}$ are given by:
\begin{equation}
	\begin{cases}
		{{{\dot{i}}}_{\text{t}x}}=\sum\limits_{k=1}^{N_{\text{TR}}}{{{{\dot{i}}}_{\text{t}x,k}}}  \\
		{{z}_{\text{t}x}}=\frac{1}{\sum\limits_{k=1}^{N_{\text{TR}}}{\frac{1}{{{z}_{\text{t}x,k}}}}}  \\
	\end{cases}
\end{equation}
where $x$ represents either 1 or 2, corresponding to TS1 and TS2, respectively. Once the parallel branches of the TS are determined, the complex power output of N-TS and A-TS can be expressed as a numerical equation in terms of ${{\delta}_{\text{A}}}$:
\begin{equation}
	\left\{
	\begin{aligned}
		{{{\bar{S}}}_{1}} &={{U}_{\text{N}}}{{\left[ \sum\limits_{i=1}^{N_{\text{TR}}+1}{a_{i}^\text{G1}{{e}^{j\left( b_{1,i}^{\text{G}1}{{\delta }_{\text{A}}}+b_{2,i}^{\text{G}1} \right)}}+c_{i}^{\text{G}1}} \right]}^{*}}  \\
		{{{\bar{S}}}_{2}} &={{U}_{\text{N}}}{{e}^{j{{\delta}_{\text{A}}}}}{{\left[ \sum\limits_{i=1}^{N_{\text{TR}}+1}{a_{i}^{\text{G}2}{{e}^{j\left( b_{1,i}^{\text{G}2}{{\delta }_{\text{A}}}+b_{2,i}^{\text{G}2} \right)}}+c_{i}^{\text{G}2}} \right]}^{*}}  \\
	\end{aligned} 
	\right.
\end{equation}
where $a_{i}^{\text{G1}}$ and $c_{i}^{\text{G1}}$ are complex numbers, $b_{1,i}^{\text{G1}}$ and $b_{2,i}^{\text{G1}}$ are real numbers. It can be observed that Eq. (6) exhibits a standard structure involving exponential polynomials. Moreover, the system equivalence process has a fully analytical expression, enabling RPAC to derive the TS output power equation through algebraic steps.

\subsection{FPAD of MSO and TSC}



For active power circulation (APC) and reactive power circulation (RPC), there are two critical conditions for the occurrence of power circulation in the system, respectively. These are when the active or reactive power of N-TS is zero (${{P}_{1}}\left( {{\delta}_{\text{A}}} \right)=0$, ${{Q}_{1}}\left( {{\delta}_{\text{A}}} \right)=0$), and when the active or reactive power of A-TS is zero (${{P}_{2}}\left( {{\delta}_{\text{A}}} \right)=0$, ${{Q}_{2}}\left( {{\delta}_{\text{A}}} \right)=0$). For example, in the case of ${{P}_{1}}\left( {{\delta }_{\text{A}}} \right)=0$, if ${{\delta }_{\text{A}}}$ is increased, ${{P}_{2}}$ will increase and ${{P}_{1}}$ will become negative, resulting in APC. The steps for obtaining FPAD of MSO and TSC are illustrated in \textbf{Algorithm 1}.

\begin{algorithm}[!t]
	\caption{The method for obatining FPAD of TSC.}\label{alg:alg1}
	\begin{algorithmic}[1]
		\REPEAT 
		\STATE \textbf{The process for obtaining} $\Xi_{\delta, \text{P}}^{\text{MSO}}$:
		\STATE Obtain $\delta_{\text{A}}^{\text{cr}}={{\arg}_{{{\delta}_{\text{A}}}\in {{\mathbf{\Delta}}_{2}}}}\left({{P}_{1}}\left( {{\delta}_{\text{A}}} \right)=0 \right)$ and\\
					$\delta_{\text{A}}^{\text{cr}}={{\arg}_{{{\delta }_{\text{A}}}\in {{\mathbf{\Delta }}_{2}}}}\left( {{P}_{2}}\left( {{\delta}_{\text{A}}} \right)=0 \right)$, respectively.
		\STATE Based on Eq. (8) and (9), obtain the FPAD corresponding to APC, $\Xi_{\delta, \text{P}}^{\text{MSO}}$.
		\STATE \textbf{The process for obtaining} $\Xi_{\delta,\text{Q}}^{\text{MSO}}$:
		\IF{Use strict RPC constraints}
		\IF{exists $\delta_{\text{A}}^{\text{cr}} = {{\arg }_{{{\delta }_{\text{A}}} \in {{\mathbf{\Delta }}_{2}}}}\left( {{Q}_{1}}\left( {{\delta}_{\text{A}}} \right) = 0 \right)$ and\\ $\delta_{\text{A}}^{\text{cr}} = {{\arg }_{{{\delta }_{\text{A}}} \in {{\mathbf{\Delta }}_{2}}}}\left( {{Q}_{2}}\left( {{\delta}_{\text{A}}} \right) = 0 \right)$}
		\STATE Calculate the bounds of FPAD using Eq. (10).
		\ELSE
		\STATE Calculate the bounds of FPAD using ${{\mathbf{\Delta}}_{\text{A}}}$ as a substitute, based on Eq. (11).
		\ENDIF
		\ELSE
		\STATE Calculate the bounds of FPAD for a given $Q_{\text{cir}}^{\text{max}}$, based on Eq. (13).
		\ENDIF
		\STATE Based on Eq. (12), obtain the FPAD corresponding to RPC, $\Xi_{\delta,\text{Q}}^{\text{MSO}}$.
		\STATE Based on Eq. (13), obtain the FPAD of current MSO with APC and RPC constraints, $\Xi_{\delta}^{\text{MSO}}$.
		\UNTIL{Both MSO1 and MSO2 are done.}
		\STATE Calculate the FPAD of TSC, $\Xi_{\delta}^{\text{TSC}}$, based on Eq. (14)
	\end{algorithmic}
	\label{alg1}
\end{algorithm}

For APC, the methods for determining the upper and lower bounds of FPAD corresponding to scenario 1 (${{P}_{1}}\left({{\delta}_{\text{A}}}\right)=0$) and scenario 2 (${{P}_{2}}\left( {{\delta}_{\text{A}}} \right)=0$) can be described as follows:
\begin{equation}
	\begin{cases}
		\delta _{P}^{\max ,\text{s}1}=\delta _{\text{A}}^{\text{cr}} & P_{2}^{\text{s1}}\ge 0 \\
		\delta _{P}^{\min ,\text{s}1}=\delta _{\text{A}}^{\text{cr}} & P_{2}^{\text{s1}}<0 \\
	\end{cases}
	\begin{cases}
		\delta _{P}^{\min ,\text{s}2}=\delta _{\text{A}}^{\text{cr}} & P_{1}^{\text{s2}}\ge 0  \\
		\delta _{P}^{\max ,\text{s}2}=\delta _{\text{A}}^{\text{cr}} & P_{1}^{\text{s2}}<0  \\
	\end{cases}
\end{equation}
where $\delta _{P}^{(\cdot),(\cdot)}$ represents the critical phase angle corresponding to APC, the symbols $\max$ and $\min$ denote the upper and lower bounds of FPAD, respectively; $\text{s1}$ and $\text{s2}$ correspond to scenario 1 and scenario 2; $P_{1}^{\text{s2}}$ and $P_{2}^{\text{s1}}$ represent the active power of N-TS and A-TS. For MSO, the FPAD corresponding to APC, denoted as $\Xi_{\delta,\text{P}}^{\text{MSO}}$, can be expressed as:
\begin{equation}
	\Xi_{\delta,\text{P}}^{\text{MSO}} = 
	\begin{cases}
		[\delta_{\text{P}}^{\min,\text{s}2}, \delta_{\text{P}}^{\max,\text{s}1}] & P_{1}^{\text{s2}} \ge 0 \vee P_{2}^{\text{s1}} \ge 0 \\
		[\delta_{\text{P}}^{\min,\text{s}1}, \delta_{\text{P}}^{\max,\text{s}2}] & P_{1}^{\text{s2}} < 0 \vee P_{2}^{\text{s1}} < 0
	\end{cases}
\end{equation}
where the symbol '$\vee$' denotes the logical OR operation.

For RPC, when $\delta _{\text{A}}^{\text{cr}}$ exists, the methods for determining the upper and lower bounds of FPAD corresponding to scenario 1 (${{Q}_{1}}\left( {{\delta }_{\text{A}}} \right)=0$) and scenario 2 (${{Q}_{2}}\left( {{\delta }_{\text{A}}} \right)=0$) can be described as follows:
\begin{equation}
	\begin{cases}
		\delta _{Q}^{\min \text{,s1}}=\delta _{\text{A}}^{\text{cr}} & Q_{2}^{\text{s1}}\ge 0  \\
		\delta _{Q}^{\max \text{,s1}}=\delta _{\text{A}}^{\text{cr}} & Q_{2}^{\text{s1}}<0  \\
	\end{cases}
	\begin{cases}
		\delta _{Q}^{\max \text{,s2}}=\delta _{\text{A}}^{\text{cr}} & Q_{1}^{\text{s2}}\ge 0  \\
		\delta _{Q}^{\min \text{,s2}}=\delta _{\text{A}}^{\text{cr}} & Q_{1}^{\text{s2}}<0  \\
	\end{cases}
\end{equation}
where $\delta _{Q}^{(\cdot),(\cdot)}$ represents the critical phase angle of RPC; $Q_{1}^{\text{s2}}$ and $Q_{2}^{\text{s1}}$ represent the reactive power of N-TS and A-TS. Under specific TR power conditions, the existence of $\delta _{\text{A}}^{\text{cr}}$ cannot be guaranteed, and the system will lack a critical phase angle constraint. However, the operation of RPAC is subject to the ${{\delta }_{\text{A}}}$ allowable adjustment limit range $\Delta _{\text{A}}=\left[ \delta _{\text{A}}^{\text{min}},\delta _{\text{A}}^{\text{max}} \right]$:
\begin{subequations}
\begin{equation}
	\begin{cases}
		\delta _{\text{A}}^{\text{cr}}=\arg {{\min }_{{{\delta }_{\text{A}}}\in {{\mathbf{\Delta }}_{\text{A}}}}}\left| {{Q}_{1}}\left( {{\delta }_{\text{A}}} \right) \right|  \\
		\left\{ \begin{matrix}
			\delta _{Q}^{\min \text{,s1}}=\delta _{\text{A}}^{\text{min}} & \delta _{\text{A}}^{\text{cr}}<0  \\
		\delta _{Q}^{\max \text{,s1}}=\delta _{\text{A}}^{\text{max}} & \delta _{\text{A}}^{\text{cr}}\ge 0  \\
		\end{matrix} \right.  \\
	\end{cases}
\end{equation}
\begin{equation}
	\begin{cases}
		\delta_{\text{A}}^{\text{cr}}=\arg {{\min }_{{{\delta }_{\text{A}}}\in {{\mathbf{\Delta}}_{\text{A}}}}}\left| {{Q}_{2}}\left( {{\delta }_{\text{A}}} \right) \right|  \\
		\left\{ \begin{matrix}
			\delta _{Q}^{\min \text{,s2}}=\delta_{\text{A}}^{\text{min}} & \delta_{\text{A}}^{\text{cr}}<0  \\
			\delta _{Q}^{\max \text{,s2}}=\delta_{\text{A}}^{\text{max}} & \delta_{\text{A}}^{\text{cr}}\ge 0  \\
		\end{matrix} \right.  \\

	\end{cases}
\end{equation}
\end{subequations}

Combining scenario 1 and scenario 2, the FPAD corresponding to RPC, denoted as $\Xi _{\delta ,\text{Q}}^{\text{MSO}}$, can be expressed as:
\begin{equation}
	\Xi_{\delta ,\text{Q}}^{\text{MSO}} = 
	\begin{cases}
		[\delta_{\text{Q}}^{\min \text{,s1}}, \delta_{\text{Q}}^{\max \text{,s2}}] & Q_{1}^{\text{s2}} \ge 0 \vee Q_{2}^{\text{s1}} \ge 0 \\
		[\delta_{\text{Q}}^{\min \text{,s2}}, \delta_{\text{Q}}^{\max \text{,s1}}] & Q_{1}^{\text{s2}} < 0 \vee Q_{2}^{\text{s1}} < 0
	\end{cases}
\end{equation}

If both APC and RPC are considered simultaneously in MSO, the FPAD can be represented as the intersection of $\Xi_{\delta ,\text{P}}^{\text{MSO}}$ and $\Xi_{\delta ,\text{Q}}^{\text{MSO}}$. Moreover, RPAC will maintain a certain phase angle adjustment margin, which defined as $\alpha_{\delta} \in [0,1]$. For MSO, the strict FPAD, $\Xi_{\delta}^{\text{MSO}}$, is represented as:
\begin{equation}
	\Xi_{\delta}^{\text{MSO}} = \alpha_{\delta} [\Xi_{\delta ,\text{P}}^{\text{MSO}} \cap \Xi_{\delta ,\text{Q}}^{\text{MSO}}]
\end{equation}


For the proposed FTPSS, the RPC constraint can be relaxed to some extent, improving system operation flexibility without significantly consuming converter capacity. The permissible RPC for the system is defined as $Q_{\text{cir}}^{\text{max}}$. Additionally, $\delta_{\text{A}}^{\text{cr}}$ can also be obtained by simply modifying the discriminant and Eq. (11) as follows:
\begin{subequations}
\begin{equation}
	\begin{cases}
		\delta_{\text{A}}^{\text{cr}} = {{\arg }_{{{\delta }_{\text{A}}} \in {{\mathbf{\Delta }}_{2}}}}\left( Q_{1}(\delta_{\text{A}}) \pm Q_{\text{cir}}^{\text{max}} = 0 \right) \\
		\delta_{\text{A}}^{\text{cr}} = {{\arg }_{{{\delta }_{\text{A}}} \in {{\mathbf{\Delta }}_{2}}}}\left( Q_{2}(\delta_{\text{A}}) \pm Q_{\text{cir}}^{\text{max}} = 0 \right) 
	\end{cases}
\end{equation}
\begin{equation}
	\Xi_{\delta ,\text{Q}}^{\text{MSO}} = 
	\begin{cases}
		[\delta_{\text{Q}}^{\min \text{,s1}}, \delta_{\text{Q}}^{\max \text{,s2}}] & Q_{1}^{\text{s2}} \ge -Q_{\text{cir}}^{\text{max}} \vee Q_{2}^{\text{s1}} \ge -Q_{\text{cir}}^{\text{max}} \\
		[\delta_{\text{Q}}^{\min \text{,s2}}, \delta_{\text{Q}}^{\max \text{,s1}}] & Q_{1}^{\text{s2}} < -Q_{\text{cir}}^{\text{max}} \vee Q_{2}^{\text{s1}} < -Q_{\text{cir}}^{\text{max}}
	\end{cases}
\end{equation}
\end{subequations}
when the system is operating in the traction state, Eq. (13) uses a positive sign, and in the braking state, a negative sign is used. Fig.3 illustrates the equivalent method for ZSO, the traction network impedance becomes $\frac{L_{\text{MSO}} Z_{0}}{2}$. Compared to the MSO, the process of calculating $\Xi_{\delta}^{\text{ZSO}}$ only involves an additional number of TRs. For TSC, the RPAC must ensure that no power circulation occurs within ZSO1 and ZSO2. The FPAD of the TSC, $\Xi_{\delta}^{\text{TSC}}$, can be expressed as:
\begin{equation}
	\Xi_{\delta}^{\text{TSC}} = \Xi_{\delta}^{\text{ZMO1}} \cap \Xi_{\delta}^{\text{ZMO2}}
\end{equation}
\begin{figure}[!t]
	\centering
	\includegraphics[width=3.3in]{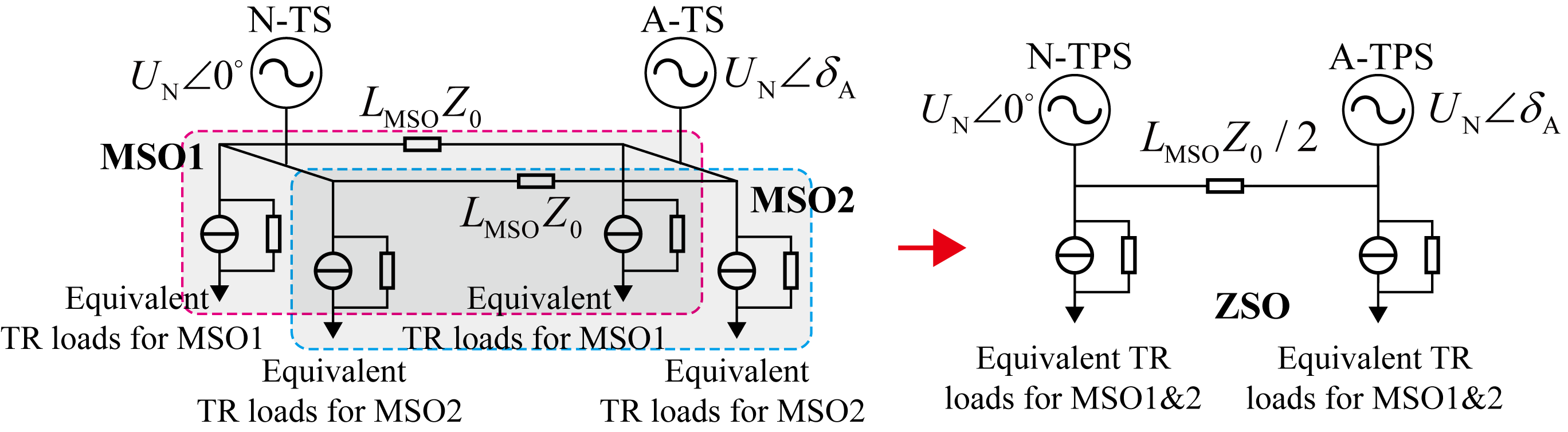}
	\caption{System equivalence method for ZSO.}
\end{figure}

\subsection{FPAD Calculation Method}
Fig.4 illustrates the flowchart for calculating $\Xi_{\delta}^{\text{TSC}}$. By separating the real and imaginary parts, Eq. (6) is transformed into a polynomial involving trigonometric functions, which can be viewed as an unconstrained single-variable numerical optimization problem. The trust region algorithm with dogleg path (TR-DP) is adopted, which ensures fast convergence and dynamic adjustment of the optimization step size and trust region radius\cite{do_nascimento_dimensioning_2022}. In the figure, typical curves of N-TS and A-TS output powers are depicted. The characteristics of these curves and the existence of zeros will be discussed in detail in the experimental section. The reactive power curve of TS may have two zero points, namely P1 and P2, or no zero point. To this end, the initial value of the algorithm is set to zero, that is $\delta_{\text{A}} = 0^{\circ}$. For the case where there are two zero points, TR-DP converges to an inner zero point P2 close to the initial value, satisfying the RPC condition. Conversely, in the absence of zero points, the TR-DP will automatically converge to the location of the marked minimum point, which ensures that the obtained $\delta_\text{A}$  minimizes the reactive TS power. In addition, we introduce the concept of FPDD to better indicate whether RPAC should adopt strict or relaxed RPC constraints. The specific definition of FPDD is provided later.
\begin{figure}[!t]
	\centering
	\includegraphics[width=3.2in]{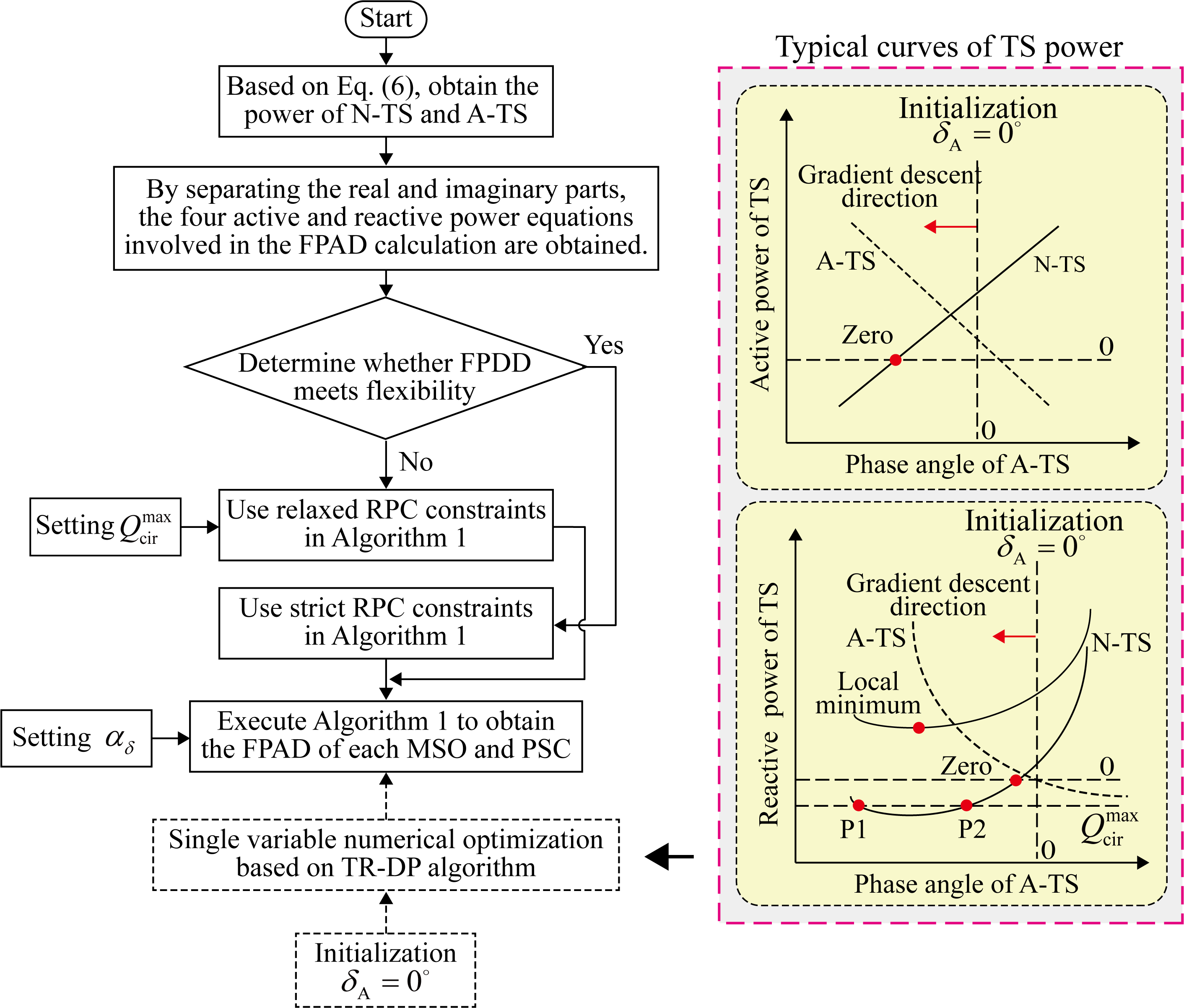}
	\caption{Flowchart for calculating FPAD of TSC, $\Xi_{\delta}^{\text{TSC}}$.}
\end{figure}

\section{Power Dispatch Method}
In this section, the concept of power allocation between TSs is defined. On this basis, the method of deriving FPDD based on FPAD is explained. Finally, within the framework of single-variable numerical optimization, the PD method with three modes are proposed.
\subsection{Definition of Power Distribution Coefficient and FPDD}
To describe the ability of A-TS and N-TS to carry TR loads, the TS power distribution coefficient $K_{\text{P}}^{\text{ZSO}}$ is defined as:
\begin{equation}
K_{\text{P}}^{\text{ZSO}} = \frac{P_{2}^{\text{ZSO}}}{P_{1}^{\text{ZSO}}}
\end{equation}
where $K_{\text{P}}^{\text{ZSO}}$ represents the ratio of the active power of A-TS, $P_{2}^{\text{ZSO}}$, to the active power of N-TS, $P_{1}^{\text{ZSO}}$. A larger \(K_{\text{P}}^{\text{ZSO}}\) indicates that A-TS will carry more of the TR loads or absorb more of BRE. Since RPAC restricts the phase angle operating range by FPAD, the system also has a maximum adjustable range for $K_{\text{P}}^{\text{ZSO}}$, referred to as the FPDD. For ZSO, we denote this by $\Xi_{\text{P}}^{\text{ZSO}}$. For a TSC, we define the $K_{\text{P}}^{\text{TSC}}$ as the ratio of the active power of A-TS, $P_{2}^{\text{TSC}}$, to the sum of the active power of the two N-TSs, $P_{1}^{\text{ZSO1}}$ and $P_{3}^{\text{ZSO2}}$:
\begin{equation}
\left\{ 
\begin{array}{l}
	K_{\text{P}}^{\text{TSC}} = \frac{P_{2}^{\text{TSC}}}{P_{1}^{\text{ZSO1}} + P_{3}^{\text{ZSO2}}} \\
	P_{2}^{\text{TSC}} = P_{2}^{\text{ZSO1}} + P_{2}^{\text{ZSO2}} \\
\end{array}
\right.
\end{equation}

The maximum adjustable range for $K_{\text{P}}^{\text{TSC}}$ denoted by $\Xi_{\text{P}}^{\text{TSC}}$.

\subsection{FPDD calculation method}
In FPAD, the function of A-TS active power output is monotonically increasing, while the function of N-TS is monotonically decreasing. Moverover, $P_1$ and $P_2$ have the same sign. For ZSO, it can be proven that $K_{\text{P}}^{\text{ZSO}}$, as a function $f_{K}^{\text{ZSO}}$ of $\delta_\text{A}$, is always monotonically increasing:
\begin{equation}
\begin{aligned}
	K_{\text{P}}^{\text{ZSO}} &= \frac{P_2(\delta_\text{A})}{P_1(\delta_\text{A})} \\
	&= f_{K}^{\text{MSO}}(\delta_\text{A}) \quad \uparrow
\end{aligned}
\end{equation}

For TSC, since the numerator $P_{2}^{\text{TSC}}$ is monotonically increasing and the denominators $P_{1}^{\text{ZSO1}}$ and $P_{3}^{\text{ZSO2}}$ are both monotonically decreasing, $K_{\text{P}}^{\text{TSC}}$ maintains the same monotonicity as ZSO and is still monotonically increasing. This implies that the upper and lower bounds of $\Xi_{\text{P}}^{\text{ZSO}}$ or $\Xi_{\text{P}}^{\text{TSC}}$ correspond to the upper and lower bounds of $\Xi_{\delta}^{\text{ZSO}}$ or $\Xi_{\delta}^{\text{TSC}}$, respectively. Therefore, for both ZSO and TSC, simply substituting the boundary of $\Xi_{\delta}^{\text{ZSO}}$ or $\Xi_{\delta}^{\text{TSC}}$ into the respective equations allows for the calculation of $\Xi_{\text{P}}^{\text{ZSO}}$ or $\Xi_{\text{P}}^{\text{TSC}}$.

\subsection{PD Method with Three Modes}
\textit{The inter-station power distribution mode (PDM)}

After receiving PD instructions of TSs from upper-level dispatch, $K_{\text{P}}^{\text{ZSO}}$ and  $K_{\text{P}}^{\text{TSC}}$ can be regarded as constants for RPAC. Thus, Eq. (16) can be transformed into:
\begin{equation}
\begin{cases}
	P_{2}^{\text{ZSO}} - K_{\text{P}}^{\text{ZSO}} P_{1}^{\text{ZSO}} = 0 \\
	P_{2}^{\text{TSC}} - K_{\text{P}}^{\text{TSC}} (P_{1}^{\text{ZSO1}} + P_{3}^{\text{ZSO2}}) = 0
\end{cases}
\end{equation}

By separating the real and imaginary parts of Eq. (6), the unified equation form of Eq. (18) can be expressed as:
\begin{equation}
\begin{aligned}
	&\sum_{i=1}^{k} \left[ \sin \left( \delta_{\text{A}} \right) + \cos ( \delta_{\text{A}} ) \right] \left[ a_{1,i} \cos \left( b_{1,i} \delta_{\text{A}} + c_{1,i} \right) + \right.\\
	& a_{2,i} \sin \left( b_{2,i} \delta_{\text{A}} + c_{2,i} \right) + d_{1,i} \left.\right]	+ \sum_{i=1}^{k} \left[ a_{3,i} \cos \left( b_{3,i} \delta_{\text{A}} + c_{3,i} \right) + \right.\\
	&a_{4,i} \sin \left( b_{4,i} \delta_{\text{A}} + c_{4,i} \right) \left.\right] + d_{2,i} = 0
\end{aligned}
\end{equation}
where $a_{x,i}, b_{x,i}, c_{x,i}, d_{x,i}$ are all real numbers. For ZSO, $k = N_{\text{TR}} + 1$. For TSC, $k = N_{\text{TR,ZSO1}} + N_{\text{TR,ZSO2}} + 2$, $N_{\text{TR,ZSO1}}$ and $N_{\text{TR,ZSO2}}$ are the numbers of TRs in ZSO1 and ZSO2, respectively. Eq. (19) remains a numerical equation in terms of $\delta_\text{A}$. Since the monotonicity of FPDD has been proven, there exists a unique solution for $\delta_\text{A}$ in Eq. (19). The same TR-DP algorithm used in the FPAD-CS can be efficiently employed to obtain the RPA in RPA-CS.

\textit{The constant power output mode (CPM)}

Given the A-TS reference power ${{P}_{\text{ref}}}$, the RPA can be obtained by solving ${{P}_{2}}\left({{\delta}_{\text{A}}} \right)={{P}_{\text{ref}}}$, which is a numerical equation with respect to $\delta_{\text{A}}$ similar to Eq. (19). Consequently, the TR-DP algorithm can efficiently solve this problem. Since the directly given ${{P}_{\text{ref}}}$ may violate the power circulation constraint, RPAC compares it with $\Xi_{\delta}^{\text{TSC}}$, and the RPA will be restricted to the range of $\Xi_{\delta }^{\text{TSC}}$.

\textit{The RES maximum consumption mode (MCM)}

In this mode, RPAC operates at the boundary of the FPAD. For systems in the traction state, RPAC running at the upper boundary of the FPAD increases the proportion of A-TS output power. In the regenerative braking state, the proportion of BRE absorbed by A-TS decreases. This allows for greater consumption of RES within A-TS. Furthermore, since RPAC tracks the FPAD boundary operation, the RPA calculation method remains consistent with the PDM.

\subsection{Performance Comparison}
\begin{table*}[htbp]
	\centering
	\renewcommand{\arraystretch}{1.5}
	\setcellgapes{1pt} 
	\makegapedcells 
	\caption{Comparison Between the Proposed Method and the Power Flow Method at Different Calculation Stages}
	\begin{tabular}{ccccccc}
		\hline
		\multirow{2}{*}{Object} & \multirow{2}{*}{Method} & \multicolumn{5}{c}{Calculation stages of RPAC} \\ \cline{3-7}
		&  &  &\makecell[c]{FPAD-CS for APC} & \makecell[c]{FPAD-CS for RPC} & \makecell[c]{FPDD-CS} & \makecell[c]{RPA-CS} \\ \hline
		\multirow{6}{*}{\makecell{MSO / \\ ZSO}} & \multirow{3}{*}{\makecell[c]{Proposed \\ method}} & Problem type & Numerical equation & Numerical equation & Numerical equation & Numerical equation \\ 
		&  & Solving times & 2 & 2 & 0 & 1 \\ 
		&  & Time complexity & \(O(T)\) & \(O(T)\) & \(O(1)\) & \(O(T)\) \\ \cline{2-7}
		& \multirow{3}{*}{\makecell[c]{Power flow \\ method}} & Problem type & PF & PF / OPF & PF & OPF \\ 
		&  & Solving times & 2 & 2$\sim$4 & 2 & 1 \\ 
		&  & Time complexity & \(O(T \cdot n^3)\) & \makecell[c]{$O(T \cdot n^3)$ / \\ $O(\log(1/\epsilon) \cdot n^3))$} & \(O(T \cdot n^3)\) & \(O(\log(1/\epsilon) \cdot n^3)\) \\ \hline
		\multirow{6}{*}{TSC} & \multirow{3}{*}{\makecell[c]{Proposed \\ method}} & Problem type & Numerical equation & Numerical equation & Numerical equation & Numerical equation \\ 
		&  & Solving times & 4 & 4 & 0 & 1 \\ 
		&  & Time complexity & \(O(T)\) & \(O(T)\) & \(O(1)\) & \(O(T)\) \\ \cline{2-7}
		& \multirow{3}{*}{\makecell[c]{Power flow \\ method}} & Problem type & PF & PF / OPF & PF & OPF \\
		&  & Solving times & 4 & 4$\sim$8 & 2 & 1 \\
		&  & Time complexity & \(O(T \cdot n^3)\) & \makecell[c]{$O(T \cdot n^3)$ / \\ $O(\log(1/\epsilon) \cdot n^3))$} & \(O(T \cdot n^3)\) & \(O(\log(1/\epsilon) \cdot n^3)\) \\ \hline
	\end{tabular}
\end{table*}

Table 1 shows the comparison between the proposed method and the power flow method at different stages. To facilitate the comparison of algorithm complexity, we assume that executing algebraic steps on an equation takes constant time $O(1)$. In the FPAD-CS and RPA-CS, since the derivative of the single-variable numerical equation can be precomputed, the time complexity of a single solution using the TR-DP is only $O(T)$, where $T$ is the number of iterations performed by the algorithm. In power system analysis, standard power flow methods include the Newton-Raphson (N-R) method and the more computationally efficient fast-decoupled method. However, considering that the high R/X ratio of traction networks does not satisfy the weak coupling condition of "$\text{P-}\theta$" and "Q-V" required for the fast-decoupled method \cite{stott_review_1974}, this paper uses the N-R method for PF to ensure sufficient solution accuracy. In N-R, techniques such as LU decomposition \cite{jankovic_application_2015}, predictor-corrector \cite{tostado_developed_2019}, and block matrix \cite{da_costa_developments_1999} have been used to improve efficiency. Considering the differences in algorithm complexity due to these detailed techniques, we use the time complexity of LU decomposition, $O(n^3)$, as a benchmark. Consequently, the complexity of PF is approximately $O(T \cdot n^3)$, where $n$ is the number of TRs operating in the system. For OPF, the complexity of the primal-dual interior-point method is approximately $O(\log(1/\epsilon) \cdot n^3)$, where $\epsilon$ is the desired accuracy \cite{wright_primal-dual_1997}. From Table I, it can be observed that the proposed method has lower algorithmic complexity and fewer solve times in all stages.

\section{Experiment and Analysis}
\subsection{Experimental Setup}
To validate the proposed methods, this study constructed an experimental platform in MATLAB. The experimental laptop CPU model is Intel i7-12700h with a base frequency of 2.3GHz. Table II presents the simulation parameters. Considering the widespread use of HXD and CR series TR in China, this paper takes the CRH2A with a maximum traction power of 4.8 MW as an example, and sets the typical operating power of TR to ${{\bar{S}}_{\text{t}}}=4\text{ MW}+j0.5\text{ MVar}$ \cite{chen_multitime-scale_2020}. 
\begin{table}[!tb]
	\centering
	\renewcommand{\arraystretch}{1.2} 
	\caption{Experimental Parameter Settings}
	\begin{tabular}{cc}
		\hline
		\makecell[c]{Experimental parameters} & \makecell[c]{Value} \\ \hline
		\makecell[c]{TR traction power $\overline{S_{\text{t}}}$} & \makecell[c]{4 MW + $j$0.5 MVAr} \\ 
		\makecell[c]{TR branch impedance $Z_{\text{t}}$} & 1 k$\Omega$ \\ 
		\makecell[c]{Traction network per \\ unit length impedance $Z_0$} & \makecell[c]{0.15 + $j$0.55 $\Omega$} \\ 
		\makecell[c]{Length of MSO / ZSO} & 40 km \\ 
		\makecell[c]{Phase angle margin $\alpha_s$} & 0.95 \\
		\makecell[c]{Allowable adjustment range \\ of phase angle $\Delta_{\text{A}}$} & \makecell[c]{[-20$^\circ$, 20$^\circ$]} \\ 
		\makecell[c]{TR speed} & \makecell[c]{300 km/h} \\ \hline
	\end{tabular}
\end{table}

\subsection{Verification of FPAD Calculation Method}
\textit{Verification of Convergence}
\begin{figure}[!tbp]
	\centering
	\subfloat[]{\includegraphics[width=2.8in]{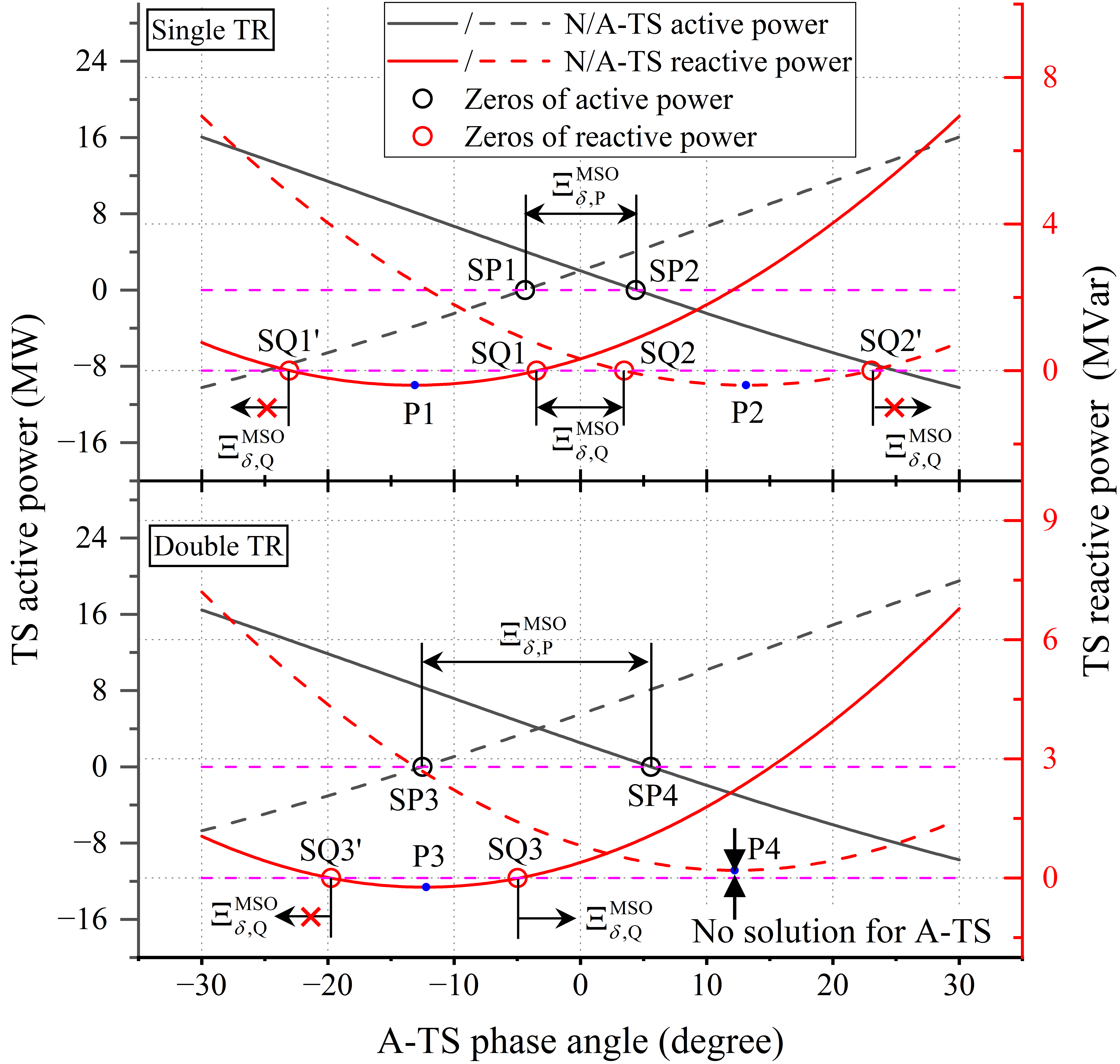}}
	\hfil
	\subfloat[]{\includegraphics[width=2.8in]{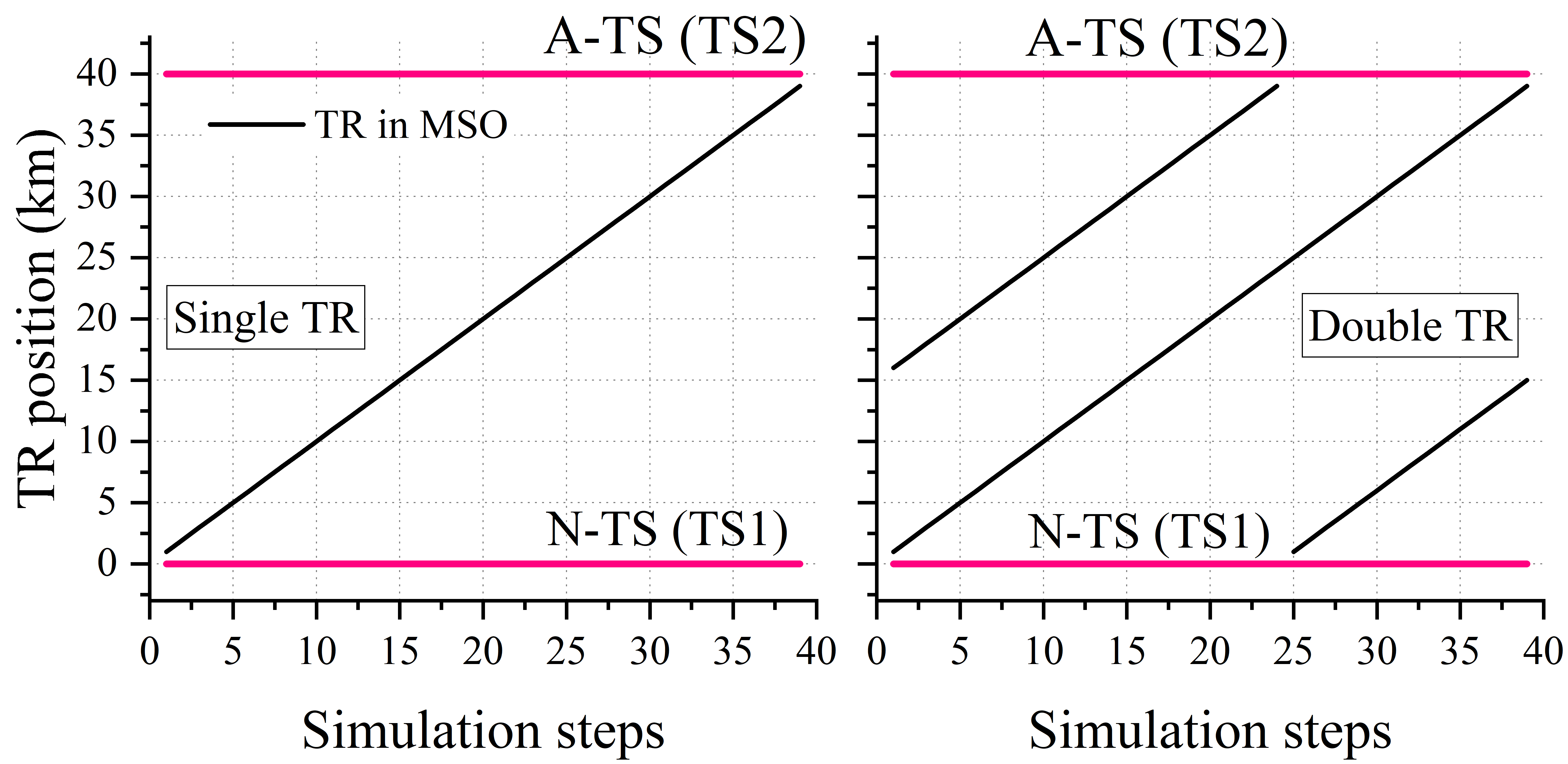}}
	\hfil
	\subfloat[]{\includegraphics[width=2.8in]{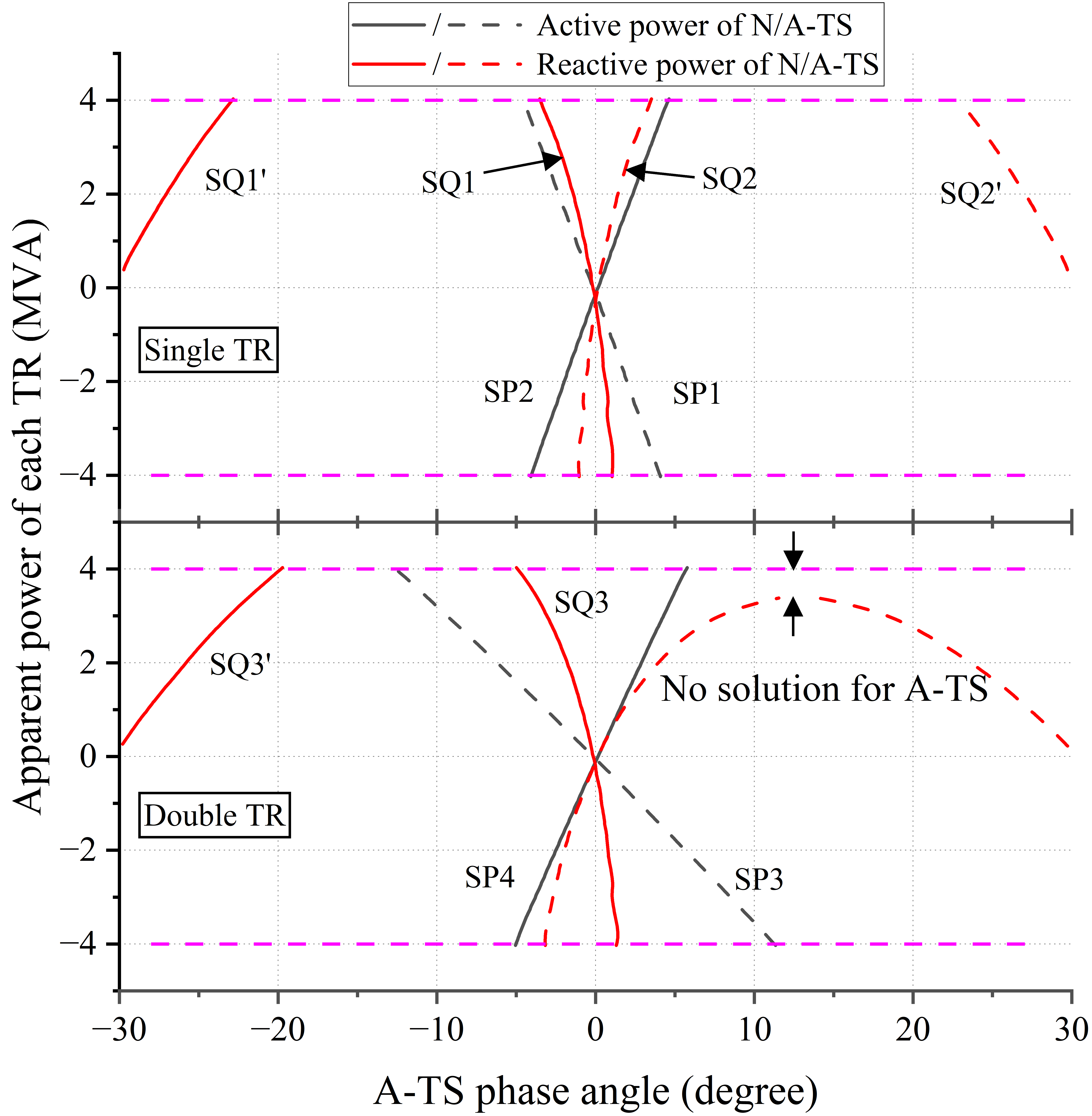}}
	\caption{Verification of FPAD calculation method. (a) The curves of N-TS and A-TS power. (b) The traffic operation schedule. (c) The zero-point trajectories of N-TS and A-TS power curves.}
\end{figure}

Fig. 5(a) shows the curves of N-TS and A-TS power when a single TR and two TRs operate in the MSO. Fig. 5(b) shows the corresponding TR scheduling plan used in the experiment. It can be observed that the curves of N-TS and A-TS active power exhibit monotonicity and have a unique zero (black circle). Based on this, the FPAD for APC, $\Xi_{\delta,\text{P}}^{\text{MSO}}$, is marked in the figure. For RPC, within ${{\mathbf{\Delta}}_{\text{A}}}$, the curves of TS reactive power are not monotonic but rather parabolic, with a minimum point (P1$\sim$P4). Additionally, there are two possible cases for the number of zeros. The first case is the existence of two distinct zeros. For instance, in the single TR experiment, three discontinuous FPAD regions of RPC, $\Xi_{\delta,\text{Q}}^{\text{MSO}}$, can be obtained. However, considering the position of $\Xi _{\delta,\text{P}}^{\text{MSO}}$, the two regions of $\Xi_{\delta,\text{Q}}^{\text{MSO}}$ far from the zero phase angle can be discarded. The second case is the absence of zeros. In the experiment for two TRs, the reactive power of TS2 is always greater than zero, and the system only contains two independent regions of $\Xi _{\delta,\text{Q}}^{\text{MSO}}$. The feasible region of $\Xi_{\delta,\text{Q}}^{\text{MSO}}$ at the right side has no upper bound. And in this case, ${{\mathbf{\Delta}}_{\text{A}}}$ constrains the phase angle. For curves with zeros, the TR-DP algorithm will gradually converge to the target zero near $\delta_{\text{A}} = 0^{\circ}$, labeled as SP1$\sim$4 and SQ1$\sim$3 in the figure, which precisely meets the criteria for $\Xi_{\delta,\text{P}}^{\text{MSO}}$ and $\Xi_{\delta,\text{Q}}^{\text{MSO}}$. Fig. 5(c) further illustrates the trajectories of zeros for different TR power levels. Negative power indicates that the TR is operating in the regenerative braking state. Fig. 5(c) shows that within ${{\mathbf{\Delta}}_{\text{A}}}$, the characteristics of zeros for the TS power remain consistent with the typical experiment, regardless of TR power adjustments. This ensures that the proposed method can stably converge to the correct RPA under various possible TR operating states.

\textit{Verification of Accuracy}
\begin{figure}[!t]
	\centering
	\includegraphics[width=2.8in]{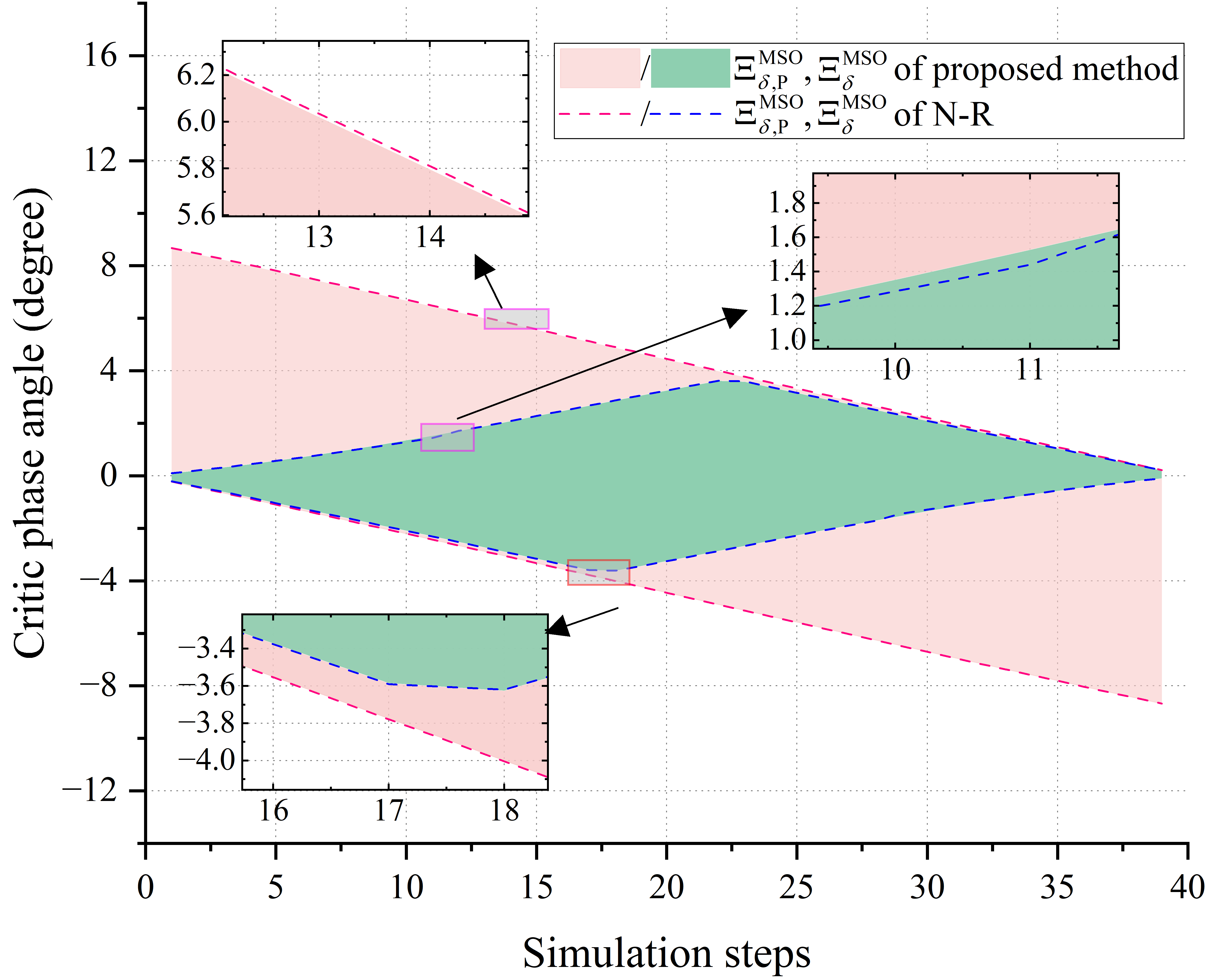}
	\caption{Comparison of FPAD results calculated by the proposed method and the N-R method.}
\end{figure}

Fig. 6 presents a comparison of FPAD results between the PM and the N-R method. The pink and green areas represent $\Xi _{\delta ,\text{P}}^{\text{MSO}}$ and $\Xi _{\delta}^{\text{MSO}}$ for the PM, respectively, while the red and green dashed lines correspond to the N-R method. It can be observed that the FPAD results of the two methods are in high agreement.

\textit{Comparison of Computing Performance}

Table III displays the time required by the proposed method and the N-R to compute $\Xi_{\delta,\text{P}}^{\text{MSO}}$ and $\Xi_{\delta,\text{Q}}^{\text{MSO}}$. In these experiments, the system base capacity is set to 100 MVA. Both methods are configured with the same calculation tolerance of 1E-8 per unit (p.u.). The results indicate that during the FPAD-CS, the proposed method achieves computational efficiency 3$\sim$4 times greater than that of the N-R. And in the RPA-CS, the proposed method's computational performance is about 10 times faster than OPF. The improvement in calculation efficiency enhances the adaptability of RPAC to uncertain TR loads.

\begin{table}[h]
	\centering
	\renewcommand{\arraystretch}{1.2}
	\caption{Comparison of Simulation Time Between the Two Methods}
	\begin{tabular}{ccccc}
		\hline
		\multirow{3}{*}{\makecell[c]{Number \\ of TRs}} & \multirow{3}{*}{Methods} &\multicolumn{3}{c}{Calculation time (ms)} \\ \cline{3-5}
		& & \multicolumn{2}{c}{\makecell[c]{FPAD-CS}} & \multirow{2}{*}{\makecell[c]{RPA-CS}} \\ \cline{3-4}
		& & APC & RPC & \\ \hline
		\multirow{2}{*}{Single} & \makecell[c]{PM} & 1.18 & 1.39 & 1.15 \\
		& \makecell[c]{PF / OPF} & 4.35 & 5.06 & 14.26 \\ \hline
		\multirow{2}{*}{Double} & \makecell[c]{PM} & 1.9 & 1.84 & 1.13 \\
		& \makecell[c]{PF / OPF} & 4.83 & 10.8 & 14.46 \\ \hline
	\end{tabular}
\end{table}

\subsection{Verification of Power Dispatch Method}
\textit{Verification within MSO}

\begin{figure}[!t]
	\centering
	\includegraphics[width=3.3in]{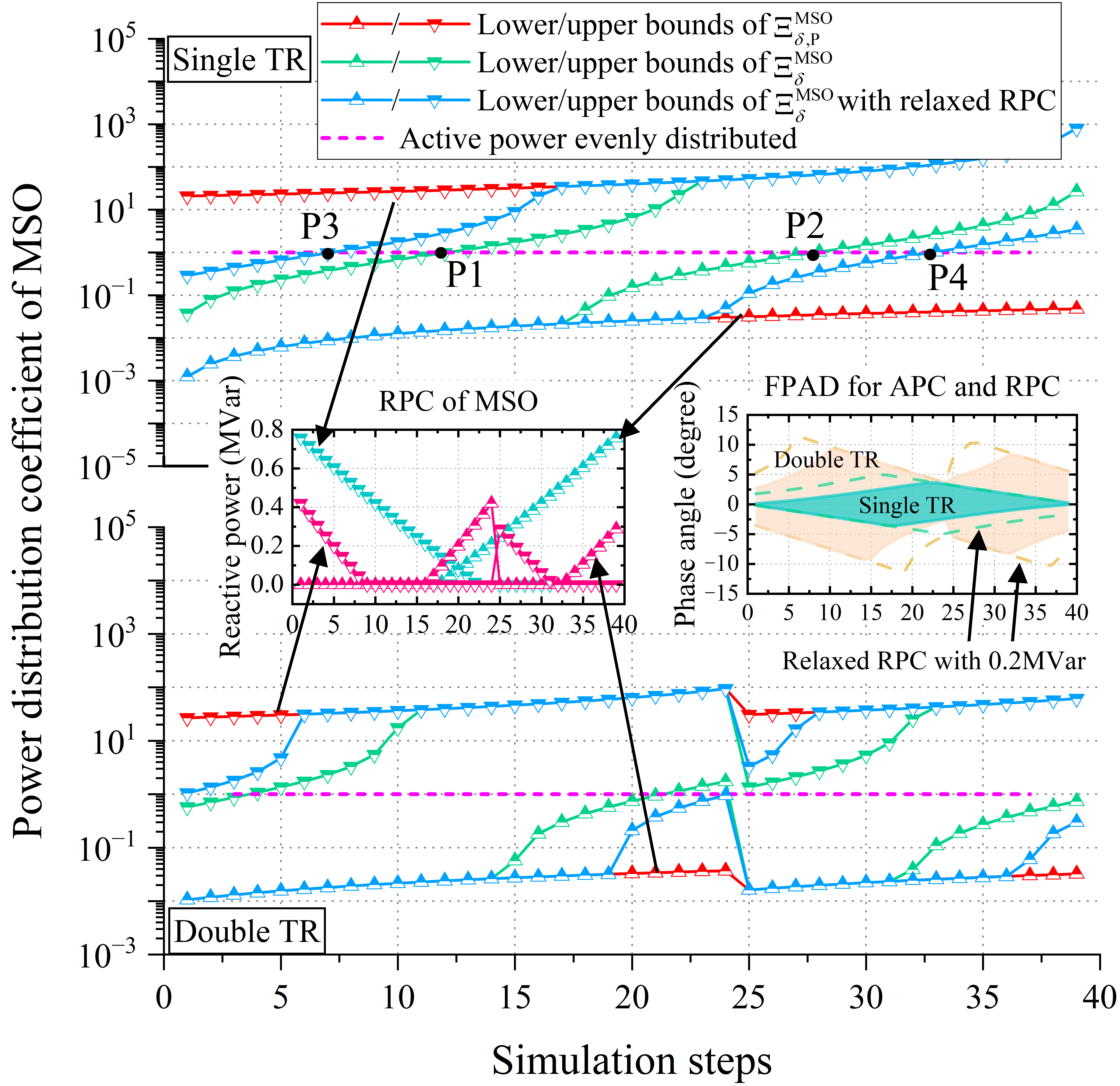}
	\caption{For an MSO operating with one or two TRs, the FPDD corresponding to FPAD and RPC.}
\end{figure}

Fig. 7 illustrates $\Xi_{\text{P}}^{\text{MSO}}$ derived from $\Xi_{\delta,\text{P}}^{\text{MSO}}$ and $\Xi_{\delta,\text{Q}}^{\text{MSO}}$ when operating one or two TRs in the MSO. Observations reveal that within the flexible FPDD region spanning from point P1 to P2, $\Xi_{\text{P}}^{\text{MSO}}$ allows adjustments where $K_{\text{P}}^{\text{MSO}}$ varies above or below 1. This flexibility implies that both N-TS and A-TS can efficiently manage traction loads without violating power circulation constraints. Moreover, the FTPSS scheduling department can flexibly allocate inter-station active power ratios based on individual RES states, thereby prioritizing RES consumption within the railway system. However, the constrained $\Xi_{\delta}^{\text{MSO}}$ (blue and pink areas in the right subplot) restricts the coverage of the flexible FPDD region.

The left subplot of Fig. 13 shows the curves of RPC obtained using $\Xi_{\delta,\text{P}}^{\text{MSO}}$ when considering only APC. The results indicate that with one TR case, RPC can reach a maximum of 0.8 MVar. With two TRs operating, RPC is lower but still exceeds 0.4 MVar. This underscores the impracticality of completely relaxing RPC constraints, as it would excessively utilize valuable power electronic converter capacity and escalate construction costs. Therefore, this paper sets an RPC tolerance $Q_{\text{cir}}^{\text{max}} = 0.2 \text{ MVar}$. The inner regions marked between the two blue lines with triangle symbols represent the new FPDD, $\Xi_{\text{P}}^{\text{MSO}}$. In this scenario, the coverage of the flexible FDD region substantially increases from 40\% (P1 to P2) to over 70\% (P3 to P4).

\textit{Verification within TSC}

Fig. 8(a) illustrates the simulated TR operation schedule, with multiple TRs operating on the up and down lines of ZSO1 and ZSO2. Fig. 8(b) shows the experimental results for PD method. The top subplot displays the system load corresponding to the experiment, including the traction state and regenerative braking state with TRs operating at full and half power. The middle subplot presents the active power of A-TS under the PDM. It can be observed that the A-TS curve resembles the shape of the total TSC load curve. This indicates that even with significant variations in TR operating states and TR power in TSC, RPAC can stably track the $K_{\text{P}}^{\text{TSC}}$ setpoint, achieving precise power distribution between A-TS and N-TS. The pink area in the phase angle subplot indicates the FPAD, $\Xi_{\delta}^{\text{TSC}}$, under relaxed RPC constraints ($Q_{\text{cir}}^{\text{max}}=0.2$ MVar). The colored dashed lines indicate the RPA corresponding to different $K_{\text{P}}^{\text{TSC}}$, all of which remain within $\Xi_{\delta}^{\text{TSC}}$ throughout the experiment, ensuring that no unexpected power circulation occurs in the system.

The bottom subplot displays the power curves of A-TS and N-TS when RPAC is executing the CPM and MCM. Within the marked CPM region, A-TS maintains a constant output of 9 MW, while the remaining traction load is carried by N-TS. This experiment demonstrates that in CPM, the proposed method can simulate traditional FTPSS by adjusting the phase angle, allowing A-TS to track the reference power. The phase angle subplot indicates the corresponding RPA curves. After the 20th simulation step, RPAC switches its operating mode, using the upper bound of $\Xi _{\delta }^{\text{TSC}}$ as the RPA. It can be observed that in the traction state, A-TS maximizes its power output within the FPAD support range, while in the braking state, A-TS minimizes the absorption of TR's BRE. Within this range, A-TS and N-TS respectively output 8.09 MWh and -1.23 MWh of electric energy, maximizing the consumption of RES within A-TS. Compared to traditional FTPSS, the system strictly ensures that no unexpected power circulation occurs throughout the process.
\begin{figure}[!t]
	\centering
	\subfloat[]{\includegraphics[width=2.2in]{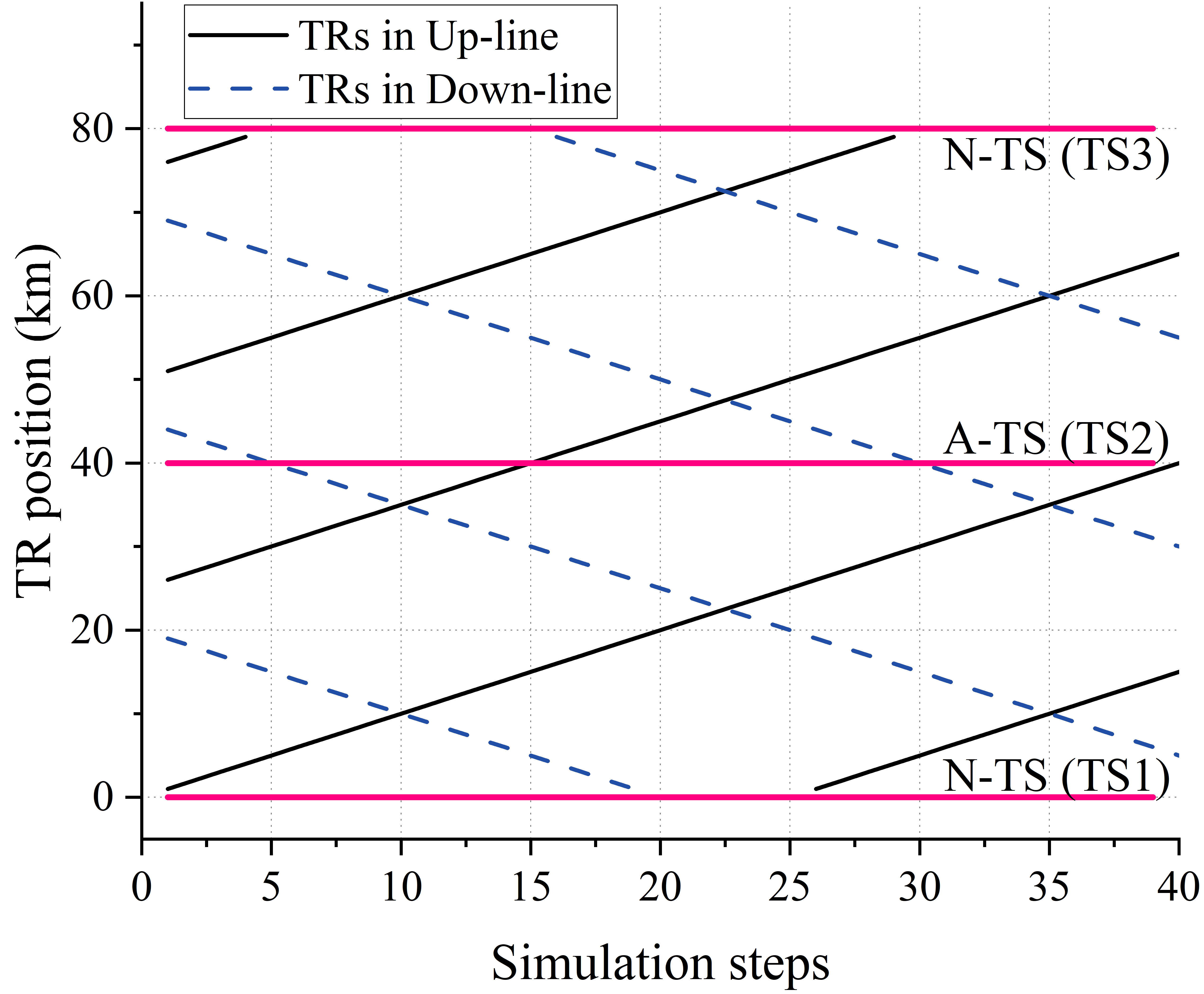}}
	\hfil
	\subfloat[]{\includegraphics[width=3.3in]{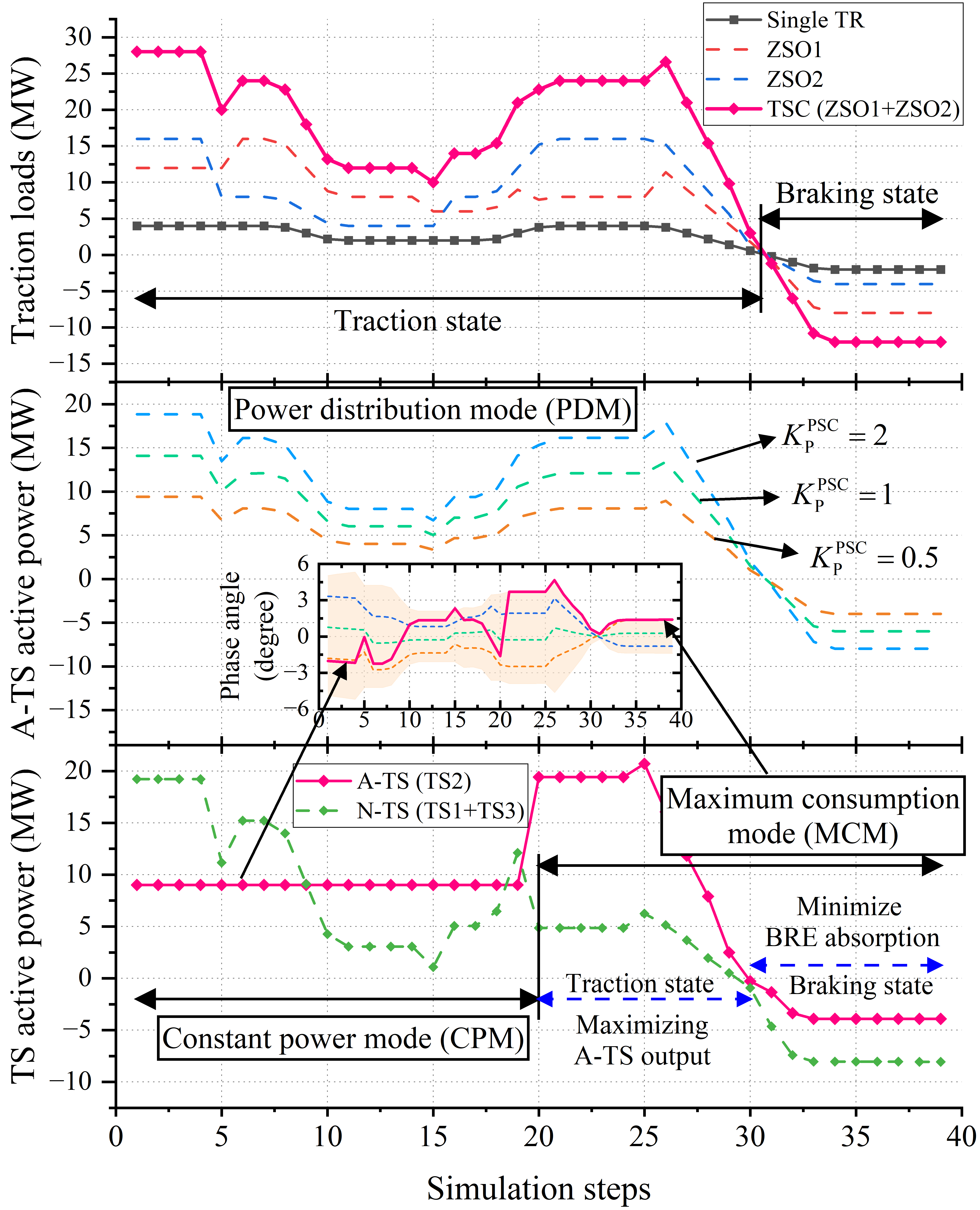}}
	\caption{Verification of PD method. (a) The TR operation schedule used in PD experiment. (b) The power distribution experiment of TSC includes PDM, CPM and MCM.}
\end{figure}

\section{Conclusion}
To address the challenges of real-time coordinated control and power mismatch in traditional FTPSS, this paper proposes a novel scheme with TSCs for PD. The TSC design, which executes local phase angle control, facilitates flexible and robust power flow management. Firstly, an equivalent model of the system with a constant topology is provided, simplifying the power flow problem to a standard form univariate numerical optimization. Based on this, feasible phase angle ranges for active and reactive power circulation are given, allowing power circulation to be strictly limited or precisely adjusted. This helps balance TS capacity utilization and PD flexibility. Secondly, three unique PD modes and the RPA calculation method based on the equivalent model are proposed for the FTPSS scheme. Under uncertain train loads, the system supports switching between TS power distribution coefficient tracking mode, constant power output mode, and maximum RES consumption mode. This flexible PD enhances the management of traction and braking power in the FTPSS. Finally, the performance of the equivalent model and PD method is validated through experiments. For field applications, the FTPSS scheme with TSCs merits further exploration, with future research findings to be summarized and published.

\bibliographystyle{IEEEtran}
\bibliography{ref}

\begin{thebibliography}{10}
\providecommand{\url}[1]{#1}
\csname url@samestyle\endcsname
\providecommand{\newblock}{\relax}
\providecommand{\bibinfo}[2]{#2}
\providecommand{\BIBentrySTDinterwordspacing}{\spaceskip=0pt\relax}
\providecommand{\BIBentryALTinterwordstretchfactor}{4}
\providecommand{\BIBentryALTinterwordspacing}{\spaceskip=\fontdimen2\font plus
\BIBentryALTinterwordstretchfactor\fontdimen3\font minus
  \fontdimen4\font\relax}
\providecommand{\BIBforeignlanguage}[2]{{%
\expandafter\ifx\csname l@#1\endcsname\relax
\typeout{** WARNING: IEEEtran.bst: No hyphenation pattern has been}%
\typeout{** loaded for the language `#1'. Using the pattern for}%
\typeout{** the default language instead.}%
\else
\language=\csname l@#1\endcsname
\fi
#2}}
\providecommand{\BIBdecl}{\relax}
\BIBdecl

\bibitem{chenVoltageUnbalanceProbability2023}
Y.~Chen, M.~Chen, M.~Cui, W.~Lu, and Y.~Lv, ``\BIBforeignlanguage{en}{Voltage
  {Unbalance} {Probability} {Pre}-{Assessment} of {Electrified} {Railways}
  {With} {Uncertain} {Traction} {Load}},'' \emph{\BIBforeignlanguage{en}{IEEE
  Trans. Transport. Electrific.}}, vol.~9, no.~1, pp. 1509--1520, Mar. 2023.

\bibitem{cheng_solar-powered_2022}
P.~Cheng, W.~Liu, J.~Ma, L.~Zhang, and L.~Jia,
  ``\BIBforeignlanguage{en}{Solar-powered rail transportation in {China}:
  {Potential}, scenario, and case},'' \emph{\BIBforeignlanguage{en}{Energy}},
  vol. 245, p. 123221, Apr. 2022.

\bibitem{yuan_optimal_2022}
J.~Yuan, K.~Cheng, and K.~Qu, ``\BIBforeignlanguage{en}{Optimal dispatching of
  high-speed railway power system based on hybrid energy storage system},''
  \emph{\BIBforeignlanguage{en}{Energy Reports}}, vol.~8, pp. 433--442, Nov.
  2022.

\bibitem{ma_rolling-adaptive_2023}
Q.~Ma, Z.~Yang, P.~Luo, Z.~Lei, and Q.~Guo, ``\BIBforeignlanguage{en}{A
  rolling-adaptive peak clipping control strategy coordinating {RBE} recycling
  and {PV} consumption},'' \emph{\BIBforeignlanguage{en}{IEEE Trans. Intell.
  Transp. Syst.}}, vol.~24, no.~4, pp. 4348--4363, Apr. 2023.

\bibitem{ge_combined_2022}
Y.~Ge, H.~Hu, J.~Chen, K.~Wang, and Z.~He, ``\BIBforeignlanguage{en}{Combined
  active and reactive power flow control strategy for flexible railway traction
  substation integrated with {ESS} and {PV}},''
  \emph{\BIBforeignlanguage{en}{IEEE Trans. Sustain. Energy}}, vol.~13, no.~4,
  pp. 1969--1981, Oct. 2022.

\bibitem{novak_hierarchical_2019}
H.~Novak, V.~Lesic, and M.~Vasak, ``\BIBforeignlanguage{en}{Hierarchical model
  predictive control for coordinated electric railway traction system energy
  management},'' \emph{\BIBforeignlanguage{en}{IEEE Trans. Intell. Transp.
  Syst.}}, vol.~20, no.~7, pp. 2715--2727, Jul. 2019.

\bibitem{chen_integrated_2023}
J.~Chen, Y.~Ge, K.~Wang, H.~Hu, Z.~He, Z.~Tian, and Y.~Li,
  ``\BIBforeignlanguage{en}{Integrated regenerative braking energy utilization
  system for multi-substations in electrified railways},''
  \emph{\BIBforeignlanguage{en}{IEEE Trans. Ind. Electron.}}, vol.~70, no.~1,
  pp. 298--310, Jan. 2023.

\bibitem{ge_hierarchical_2023}
Y.~Ge, H.~Hu, J.~Chen, K.~Wang, and Z.~He,
  ``\BIBforeignlanguage{en}{Hierarchical energy management of networked
  flexible traction substations for efficient {RBE} and {PV} energy utilization
  within {ERs}},'' \emph{\BIBforeignlanguage{en}{IEEE Trans. Sustain. Energy}},
  vol.~14, no.~3, pp. 1397--1410, Jul. 2023.

\bibitem{huang_joint_2024}
Y.~Huang, H.~Hu, Y.~Ge, H.~Liao, J.~Luo, S.~Gao, and Z.~He,
  ``\BIBforeignlanguage{en}{Joint sizing optimization method of {PVs}, hybrid
  energy storage systems, and power flow controllers for flexible traction
  substations in electric railways},'' \emph{\BIBforeignlanguage{en}{IEEE
  Trans. Sustain. Energy}}, vol.~15, no.~2, pp. 1210--1223, Apr. 2024.

\bibitem{wangPowerAllocationStrategy2023}
X.~Wang, Y.~Luo, B.~Qin, and L.~Guo, ``\BIBforeignlanguage{en}{Power
  {Allocation} {Strategy} for {Urban} {Rail} {HESS} {Based} on {Deep}
  {Reinforcement} {Learning} {Sequential} {Decision} {Optimization}},''
  \emph{\BIBforeignlanguage{en}{IEEE Trans. Transport. Electrific.}}, vol.~9,
  no.~2, pp. 2693--2710, Jun. 2023.

\bibitem{zhang_analysis_2023}
J.~Zhang, G.~Wang, G.~Feng, and Y.~Li, ``\BIBforeignlanguage{en}{Analysis of
  the impact of traction power supply system containing new energy on the power
  quality of the power system},'' \emph{\BIBforeignlanguage{en}{Energy
  Reports}}, vol.~9, pp. 363--371, Sep. 2023.

\bibitem{feng_evaluation_2020}
D.~Feng, H.~Zhu, X.~Sun, and S.~Lin, ``\BIBforeignlanguage{en}{Evaluation of
  power supply capability and quality for traction power supply system
  considering the access of distributed generations},''
  \emph{\BIBforeignlanguage{en}{IET Renew. Power Gener.}}, vol.~14, no.~18, pp.
  3644--3652, 2020.

\bibitem{zahedmanesh_sequential_2021}
A.~Zahedmanesh, K.~M. Muttaqi, and D.~Sutanto, ``\BIBforeignlanguage{en}{A
  sequential decision-making process for optimal technoeconomic operation of a
  grid-connected electrical traction substation integrated with solar {PV} and
  {BESS}},'' \emph{\BIBforeignlanguage{en}{IEEE Trans. Ind. Electron.}},
  vol.~68, no.~2, pp. 1353--1364, Feb. 2021.

\bibitem{lin_energy_2023}
J.~Lin, S.~Lin, Y.~Li, S.~Hu, J.~Zhang, J.~Zhang, B.~An, Z.~Zhang, B.~Xie,
  F.~Zhou, Y.~Cao, and J.~Yu, ``\BIBforeignlanguage{en}{An energy storage-based
  railway power flow controller with partial compensation strategy},''
  \emph{\BIBforeignlanguage{en}{IEEE Trans. Ind. Appl.}}, vol.~59, no.~1, pp.
  1222--1234, Jan. 2023.

\bibitem{chen_bi-hierarchy_2023}
M.~Chen, X.~Gong, Z.~Liang, J.~Zhao, and Z.~Tian, ``Bi-hierarchy capacity
  programming of co-phase {TPSS} with {PV} and {HESS} for minimum life cycle
  cost,'' \emph{Int. J. Electr. Power Energy Syst.}, vol. 147, p. 108904, May
  2023.

\bibitem{chen_configuration_2023}
M.~Chen, X.~Dai, J.~Lai, Y.~Chen, S.~Hillmansen, and Z.~Tian, ``Configuration
  and control strategy of flexible traction power supply system integrated with
  energy storage and photovoltaic,'' \emph{Int. J. Electr. Power Energy Syst.},
  vol. 153, p. 109410, Nov. 2023.

\bibitem{chen_chance-constrained_2024}
Y.~Chen, M.~Chen, L.~Xu, and Z.~Liang,
  ``\BIBforeignlanguage{en}{Chance-constrained optimization of storage and
  {PFC} capacity for railway electrical smart grids considering uncertain
  traction load},'' \emph{\BIBforeignlanguage{en}{IEEE Trans. Smart Grid}},
  vol.~15, no.~1, pp. 286--298, Jan. 2024.

\bibitem{liu_robust_2022}
Y.~Liu, M.~Chen, Z.~Cheng, Y.~Chen, and Q.~Li, ``\BIBforeignlanguage{en}{Robust
  energy management of high-speed railway co-phase traction substation with
  uncertain {PV} generation and traction load},''
  \emph{\BIBforeignlanguage{en}{IEEE Trans. Intell. Transp. Syst.}}, vol.~23,
  no.~6, pp. 5079--5091, Jun. 2022.

\bibitem{zhang_research_2022}
L.~Zhang, X.~Li, S.~Liang, and D.~Han, ``\BIBforeignlanguage{en}{Research on
  the influence of electric railway bilateral power supply on power system and
  countermeasures},'' \emph{\BIBforeignlanguage{en}{Int. J. Electr. Power
  Energy Syst.}}, vol. 137, p. 107769, May 2022.

\bibitem{chen_dynamic_2022}
Y.~Chen, M.~Chen, Z.~Liang, and L.~Liu, ``\BIBforeignlanguage{en}{Dynamic
  voltage unbalance constrained economic dispatch for electrified railways
  integrated energy storage},'' \emph{\BIBforeignlanguage{en}{IEEE Trans. Ind.
  Informat.}}, vol.~18, no.~11, pp. 8225--8235, Nov. 2022.

\bibitem{chen_unified_2022}
M.~Chen, Y.~Chen, Y.~Chen, X.~Dai, and L.~Liu,
  ``\BIBforeignlanguage{en}{Unified power quality management for traction
  substation groups connected to weak power grids},''
  \emph{\BIBforeignlanguage{en}{IEEE Trans. Power Del.}}, vol.~37, no.~5, pp.
  4178--4189, Oct. 2022.

\bibitem{chen_modelling_2021}
L.~Chen, M.~Chen, Y.~Chen, Y.~Chen, Y.~Cheng, and N.~Zhao,
  ``\BIBforeignlanguage{en}{Modelling and control of a novel {AT}-fed co-phase
  traction power supply system for electrified railway},''
  \emph{\BIBforeignlanguage{en}{Int. J. Electr. Power Energy Syst.}}, vol. 125,
  p. 106405, Feb. 2021.

\bibitem{liu_co-phase_2020}
L.~Liu, N.~Dai, K.~W. Lao, and W.~Hua, ``\BIBforeignlanguage{en}{A co-phase
  traction power supply system based on asymmetric three-leg hybrid power
  quality conditioner},'' \emph{\BIBforeignlanguage{en}{IEEE Trans. Veh.
  Technol.}}, vol.~69, no.~12, pp. 14\,645--14\,656, Dec. 2020.

\bibitem{ying_online_2023}
Y.~Ying, Q.~Liu, M.~Wu, and Y.~Zhai, ``\BIBforeignlanguage{en}{Online energy
  management strategy of the flexible smart traction power supply system},''
  \emph{\BIBforeignlanguage{en}{IEEE Trans. Transport. Electrific.}}, vol.~9,
  no.~1, pp. 981--994, Mar. 2023.

\bibitem{chen_optimal_2022}
M.~Chen, Z.~Liang, Z.~Cheng, J.~Zhao, and Z.~Tian,
  ``\BIBforeignlanguage{en}{Optimal scheduling of {FTPSS} with {PV} and {HESS}
  considering the online degradation of battery capacity},''
  \emph{\BIBforeignlanguage{en}{IEEE Trans. Transport. Electrific.}}, vol.~8,
  no.~1, pp. 936--947, Mar. 2022.

\bibitem{chen_multitime-scale_2020}
M.~Chen, Z.~Cheng, Y.~Liu, Y.~Cheng, and Z.~Tian,
  ``\BIBforeignlanguage{en}{Multitime-scale optimal dispatch of railway {FTPSS}
  based on model predictive control},'' \emph{\BIBforeignlanguage{en}{IEEE
  Trans. Transport. Electrific.}}, vol.~6, no.~2, pp. 808--820, Jun. 2020.

\bibitem{li_modeling_2023}
J.~Li, Y.~Wei, X.~Li, C.~Lu, X.~Guo, Y.~Lin, and M.~Molinas,
  ``\BIBforeignlanguage{en}{Modeling and stability prediction for the
  static-power-converters interfaced flexible {AC} traction power supply system
  with power sharing scheme},'' \emph{\BIBforeignlanguage{en}{Int. J. Electr.
  Power Energy Syst.}}, vol. 154, p. 109401, Dec. 2023.

\bibitem{vazquezFullyDecentralizedAdaptive2019}
N.~Vazquez, S.~S. Yu, T.~K. Chau, T.~Fernando, and H.~H.-C. Iu,
  ``\BIBforeignlanguage{en}{A {Fully} {Decentralized} {Adaptive} {Droop}
  {Optimization} {Strategy} for {Power} {Loss} {Minimization} in {Microgrids}
  {With} {PV}-{BESS}},'' \emph{\BIBforeignlanguage{en}{IEEE Trans. Energy
  Convers.}}, vol.~34, no.~1, pp. 385--395, Mar. 2019.

\bibitem{qiuPulsarCalibratedTimingSource2022}
W.~Qiu, H.~Yin, L.~Zhang, X.~Luo, W.~Yao, L.~Zhu, and Y.~Liu,
  ``\BIBforeignlanguage{en}{Pulsar-{Calibrated} {Timing} {Source} for
  {Synchronized} {Sampling}},'' \emph{\BIBforeignlanguage{en}{IEEE Trans. Smart
  Grid}}, vol.~13, no.~2, pp. 1654--1657, Mar. 2022.

\bibitem{phadke_phasor_2018}
A.~G. Phadke and T.~Bi, ``\BIBforeignlanguage{en}{Phasor measurement units,
  {WAMS}, and their applications in protection and control of power systems},''
  \emph{\BIBforeignlanguage{en}{J. Mod Power Syst. Clean Energy}}, vol.~6,
  no.~4, pp. 619--629, Jul. 2018.

\bibitem{do_nascimento_dimensioning_2022}
V.~F. Do~Nascimento, I.~Yahyaoui, R.~Fiorotti, A.~E. Amorim, I.~C. Belisário,
  C.~E. Abreu, H.~R. Rocha, and F.~Tadeo,
  ``\BIBforeignlanguage{en}{Dimensioning and efficiency evaluation of a hybrid
  photovoltaic thermal system in a tropical climate region},''
  \emph{\BIBforeignlanguage{en}{Sustainable Energy, Grids and Networks}},
  vol.~32, p. 100954, Dec. 2022.

\bibitem{stott_review_1974}
B.~Stott, ``\BIBforeignlanguage{en}{Review of load-flow calculation methods},''
  \emph{\BIBforeignlanguage{en}{Proc. IEEE}}, vol.~62, no.~7, pp. 916--929,
  1974.

\bibitem{jankovic_application_2015}
S.~Janković and B.~Ivanović, ``\BIBforeignlanguage{en}{Application of
  combined {Newton}–{Raphson} method to large load flow models},''
  \emph{\BIBforeignlanguage{en}{Electr. Pow. Syst. Res.}}, vol. 127, pp.
  134--140, Oct. 2015.

\bibitem{tostado_developed_2019}
M.~Tostado, S.~Kamel, and F.~Jurado, ``\BIBforeignlanguage{en}{Developed
  {Newton}-{Raphson} based {Predictor}-{Corrector} load flow approach with high
  convergence rate},'' \emph{\BIBforeignlanguage{en}{Int. J. Electr. Power
  Energy Syst.}}, vol. 105, pp. 785--792, Feb. 2019.

\bibitem{da_costa_developments_1999}
V.~Da~Costa, N.~Martins, and J.~Pereira, ``\BIBforeignlanguage{en}{Developments
  in the {Newton} {Raphson} power flow formulation based on current
  injections},'' \emph{\BIBforeignlanguage{en}{IEEE Trans. Power Syst.}},
  vol.~14, no.~4, pp. 1320--1326, Nov. 1999.

\bibitem{wright_primal-dual_1997}
S.~J. Wright, \emph{Primal-dual interior-point methods}.\hskip 1em plus 0.5em
  minus 0.4em\relax Philadelphia: Society for Industrial and Applied
  Mathematics, 1997.

\end{thebibliography}


\vfill

\end{document}